\documentclass[%
 reprint,
 amsmath,amssymb,
 aps,
]{revtex4-2}

\usepackage{graphicx}
\usepackage{dcolumn}
\usepackage{bm}

\usepackage{xcolor}
\newcommand{\jl}[1]{\textcolor{black}{#1}}

\begin{document}

\preprint{APS/123-QED}

\title{Piecewise omnigenous magnetohydrodynamic equilibria as fusion reactor candidates}

\author{V. Fern\'andez-Pacheco}
\affiliation{Laboratorio Nacional de Fusi\'on, CIEMAT, 28040 Madrid, Spain}
\affiliation{Universidad Carlos III, Madrid, Spain}

\author{J.L. Velasco}
\email{joseluis.velasco@ciemat.es}
\affiliation{Laboratorio Nacional de Fusi\'on, CIEMAT, 28040 Madrid, Spain}

\author{E. S\'anchez}
\affiliation{Laboratorio Nacional de Fusi\'on, CIEMAT, 28040 Madrid, Spain}

\author{R. Gaur}
\affiliation{Princeton University, New Jersey, USA}

\author{J.M. Garc\'ia-Rega\~na}
\affiliation{Laboratorio Nacional de Fusi\'on, CIEMAT, 28040 Madrid, Spain}

\author{J.A. Alonso}
\affiliation{Laboratorio Nacional de Fusi\'on, CIEMAT, 28040 Madrid, Spain}

\author{I. Calvo}
\affiliation{Laboratorio Nacional de Fusi\'on, CIEMAT, 28040 Madrid, Spain}

\author{D. Carralero}
\affiliation{Laboratorio Nacional de Fusi\'on, CIEMAT, 28040 Madrid, Spain}

\date{\today}

\begin{abstract}

In piecewise omnigenous magnetic fields, charged particles remain perfectly confined in the absence of collisions and turbulence. This concept extends the traditional notion of omnigenity, the theoretical principle upon which most of existing magnetic fusion reactor designs, including tokamaks, are based. While piecewise omnigenity broadens the range of potentially viable stellarator reactor candidates, it is achieved by relaxing the requirement of continuity in the magnetic field strength, which could appear to pose significant challenges for the design of magnetohydrodynamic equilibria. In this work, a stellarator magnetic configuration is presented that satisfies the ideal magnetohydrodynamic equilibrium equation and that achieves unprecedented levels of piecewise omnigenity. As a result, it exhibits favorable transport characteristics, including reduced bulk radial (neoclassical and turbulent) transport, bootstrap current and fast ion losses. In addition, the configuration displays robust MHD stability across a range of $\beta$ values and possesses a rotational transform profile compatible with an island divertor. Collectively, these features satisfy the standard set of physics criteria required for a viable reactor candidate which, until now, were believed to be attainable only by certain types of omnigenous stellarators.

\end{abstract}

\maketitle

\begin{figure*}[ht]
\includegraphics[angle=0,width=2\columnwidth]{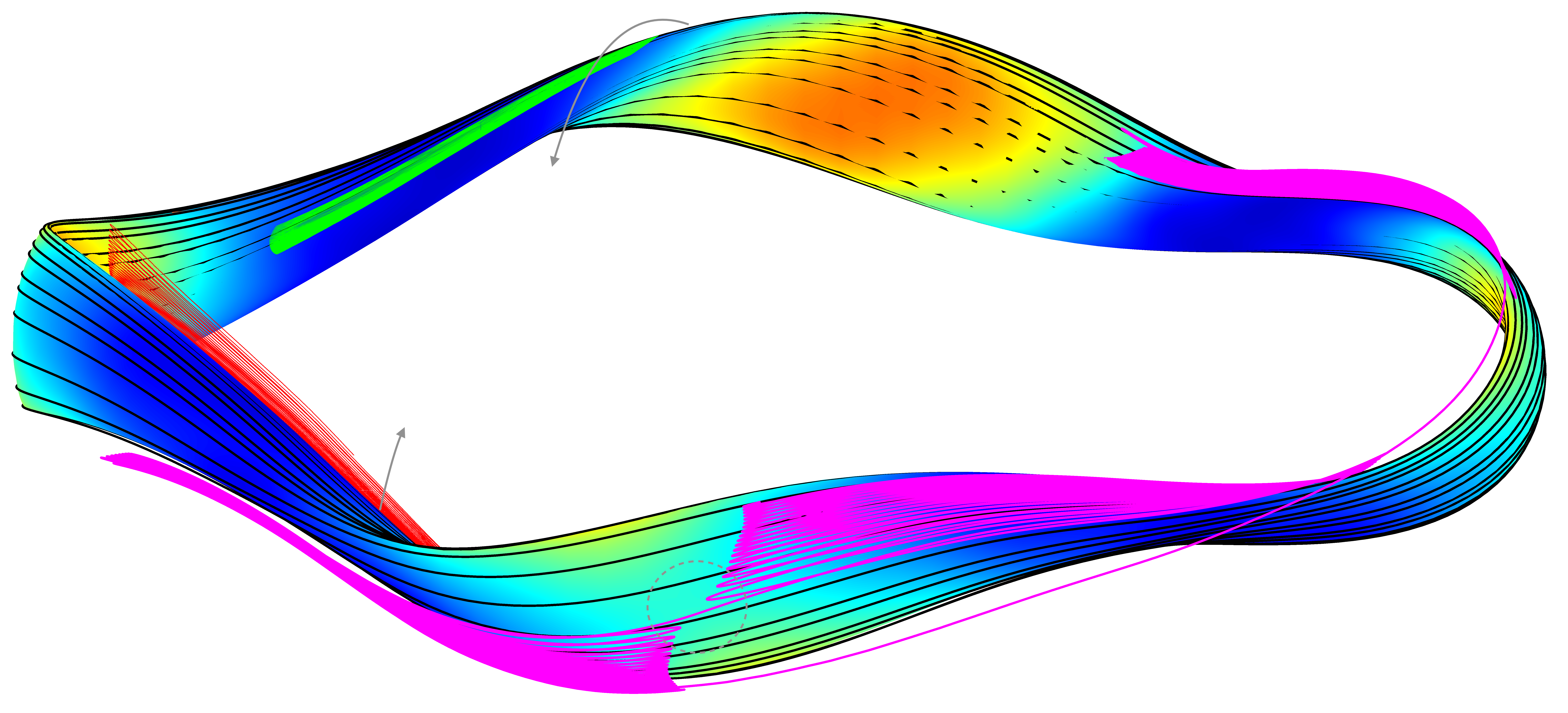}
\caption{Flux surface of a stellarator magnetic configuration (red/blue colours correspond to a larger/smaller $B$) with the guiding-center trajectories of a passing particle (black), an unconfined trapped particle (red), a confined trapped particle (green), and a transitioning particle (pink). Gray arrows represent the orbit-averaged drifts and the dashed circle highlights a transition: at some point of the trajectory of a trapped particle, one of the bounce points lies on a local maximum of $B$; immediately after this happens, the particle transitions to a different region of the magnetic surface.\label{FIG_ORBITS}}
\end{figure*}

\section{Introduction}\label{SEC_INTRO}

The two leading concepts in magnetic confinement fusion, tokamaks and stellarators~\cite{helander2012stell}, confine the charged particles of a fusion plasma~\cite{lawson1957criteria} by means of magnetic fields in which the field lines form nested toroidal flux surfaces, see e.g. figure \ref{FIG_ORBITS}. Charged particles undergo rapid gyromotion about magnetic field lines, characterized by a gyrofrequency and an associated gyroradius~\cite{northrop1961gc}. To lowest order in an expansion in the gyroperiod (the inverse of the gyrofrequency), the orbit centers, known as guiding centers, follow the magnetic field lines. On time scales much longer than the gyroperiod, the guiding-centers drift perpendicularly to the magnetic field $\mathbf{B}$, in the directions tangential and perpendicular to the flux surfaces. Cross-surface (also termed 'radial') transport of energy limits the performance of a fusion reactor, and its minimization is a fundamental design criterion. One of the main channels through which energy and particles are lost is neoclassical transport, associated to the combination of radial drifts and collisions \cite{helander2002collisional}. Another important channel is turbulent transport, which arises when small-scale instabilities in the plasma generate fluctuations in the electric and magnetic fields that modify particle orbits.

According to their lowest-order orbits, particles in a magnetic confinement device can be classified into two groups: passing and trapped particles, depicted in figure \ref{FIG_ORBITS}. For passing particles (figure \ref{FIG_ORBITS}, black), the component of the velocity that is parallel to $\mathbf{B}$ never vanishes, and they go over the complete flux surface. In the absence of collisions and turbulence, passing particles are always well confined because their radial drift velocity vanishes when averaged over lowest-order orbits. The situation is in general different for trapped particles (figure \ref{FIG_ORBITS}, green and red). For trapped particles, the component of the velocity that is parallel to $\mathbf{B}$ vanishes at two points of the trajectory called 'bounce points', where it changes sign (at these points, the value of the magnetic field strength, $B=|\mathbf{B}|$, equals the ratio between the particle energy and the particle magnetic moment $\mathcal{E}/\mu$). Trapped particles move back and forth along the magnetic field line between the bounce points while slowly drifting perpendicularly to $\mathbf{B}$. Depending on whether their radial drift velocity vanishes when averaged over lowest-order orbits (figure \ref{FIG_ORBITS}, green) or remains finite (figure \ref{FIG_ORBITS}, red), they are confined (in the absence of collisions) or they quickly leave the device, respectively. Transitioning particles (a subclass of trapped particles \cite{cary1986transitions}, figure \ref{FIG_ORBITS}, pink, \jl{see also below}) \jl{visit different regions of a flux surface and may thus} spend part of their trajectory moving on the flux surface and part drifting away from it. 

\begin{figure}
\includegraphics[angle=0,width=0.49\columnwidth]{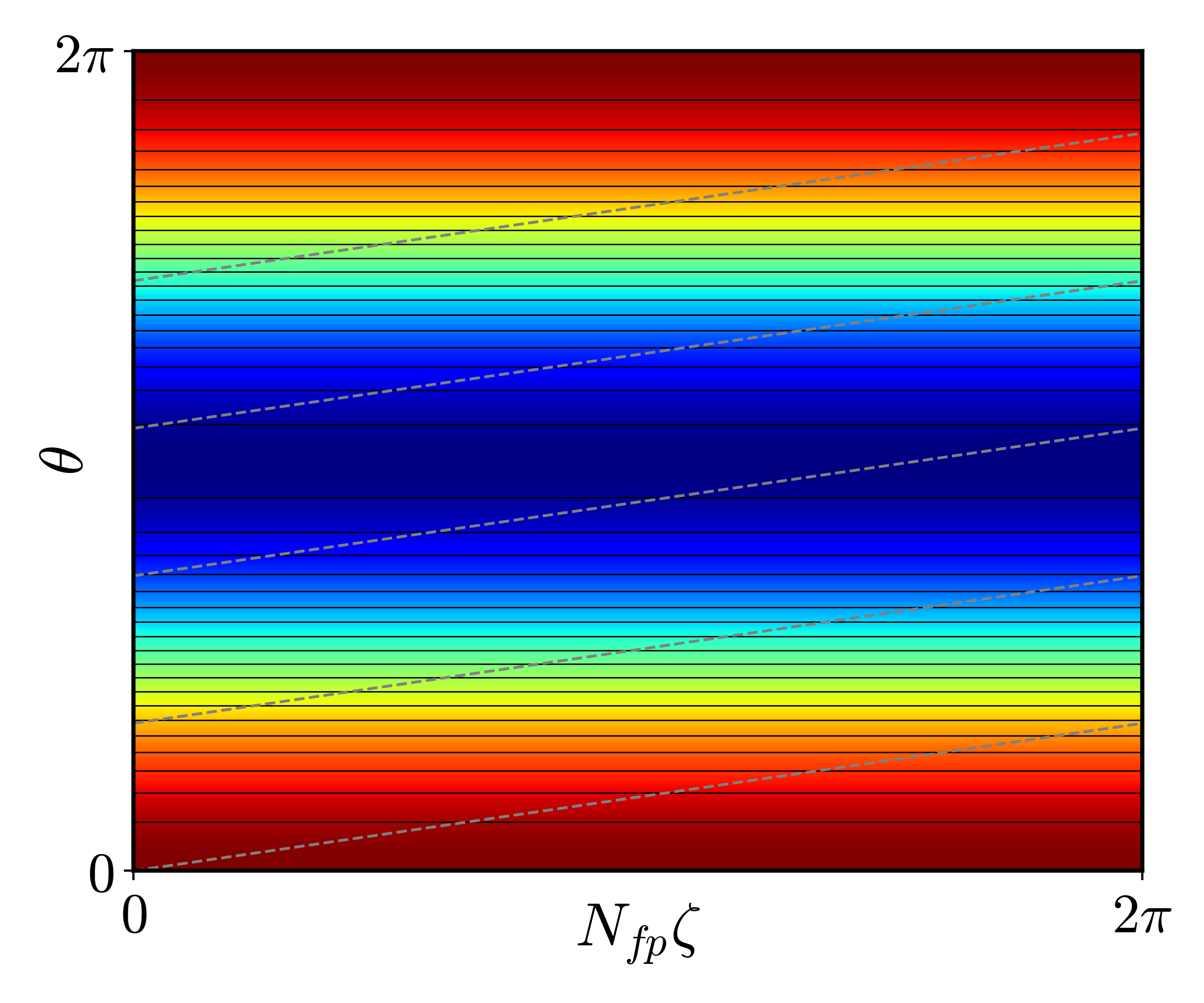}
\includegraphics[angle=0,width=0.49\columnwidth]{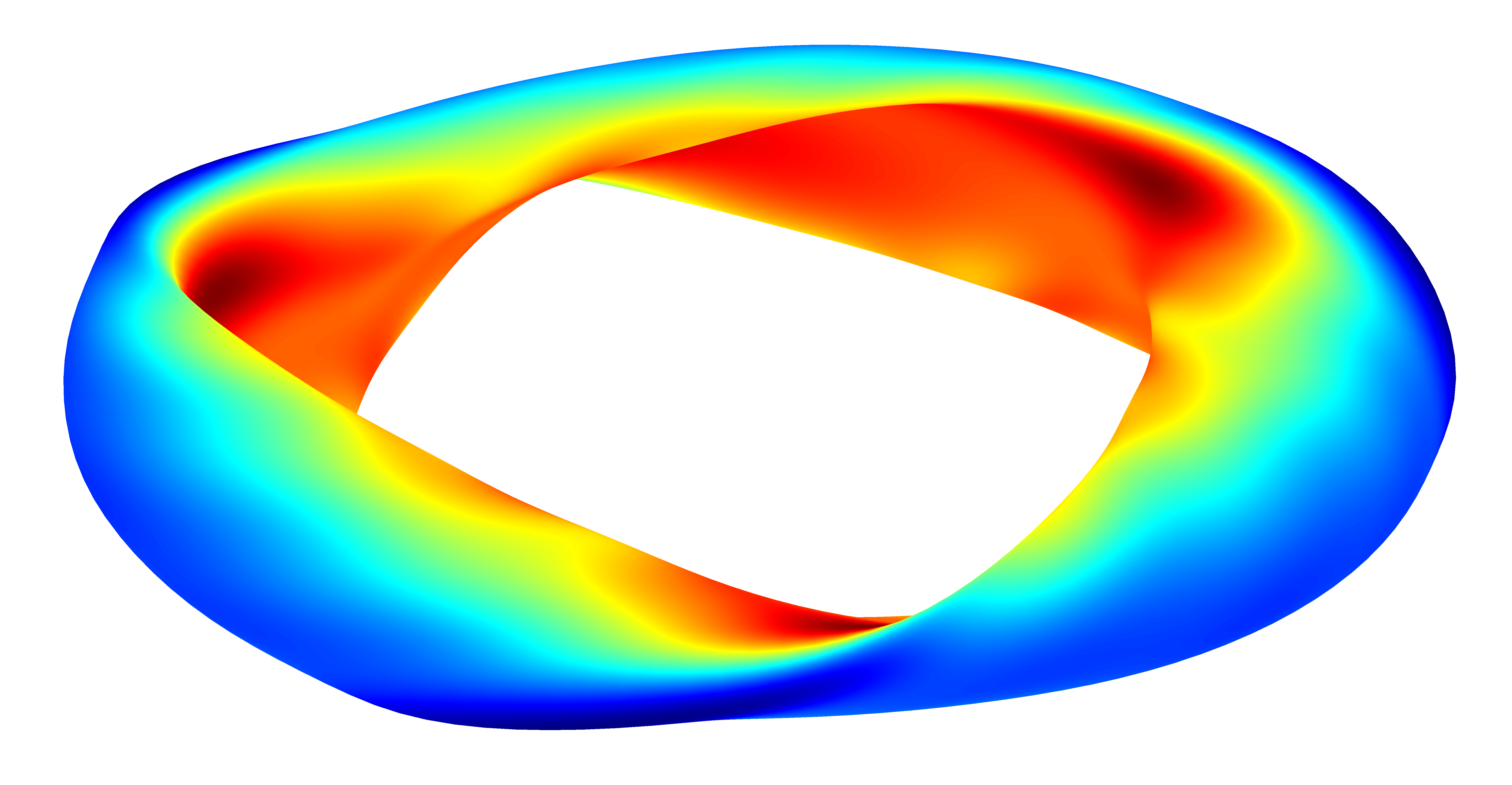}
\includegraphics[angle=0,width=0.49\columnwidth]{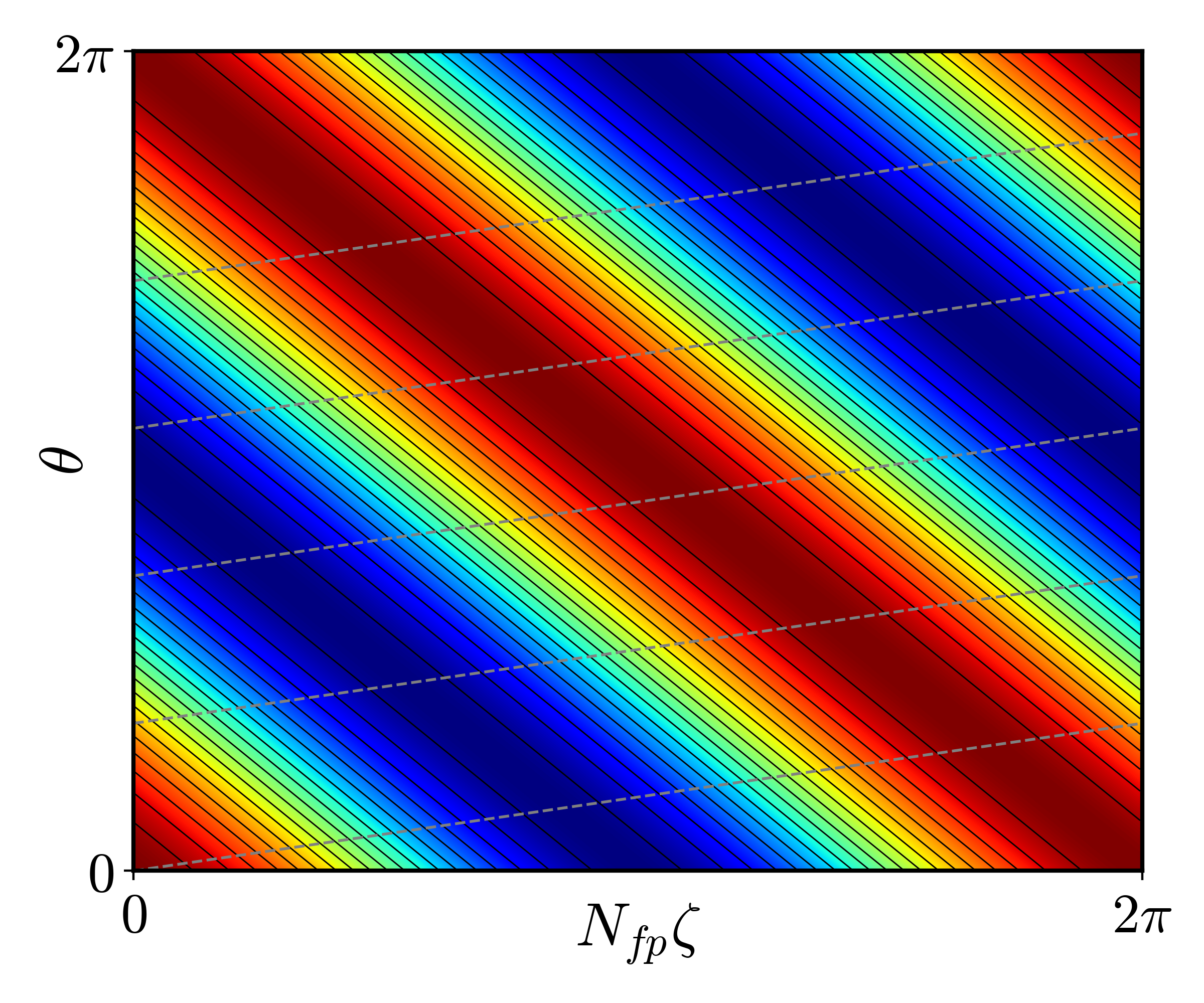}
\includegraphics[angle=0,width=0.49\columnwidth]{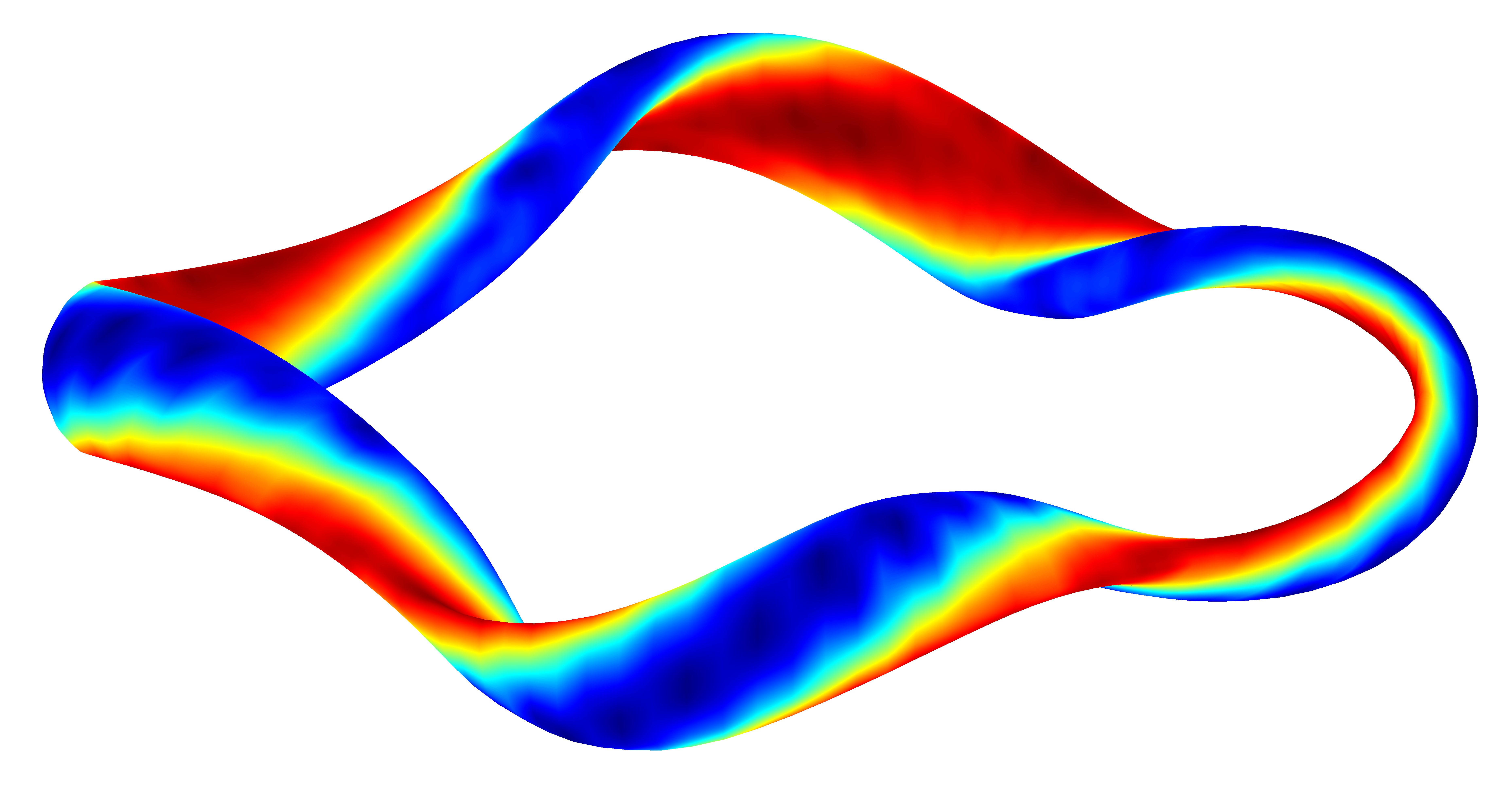}
\includegraphics[angle=0,width=0.49\columnwidth]{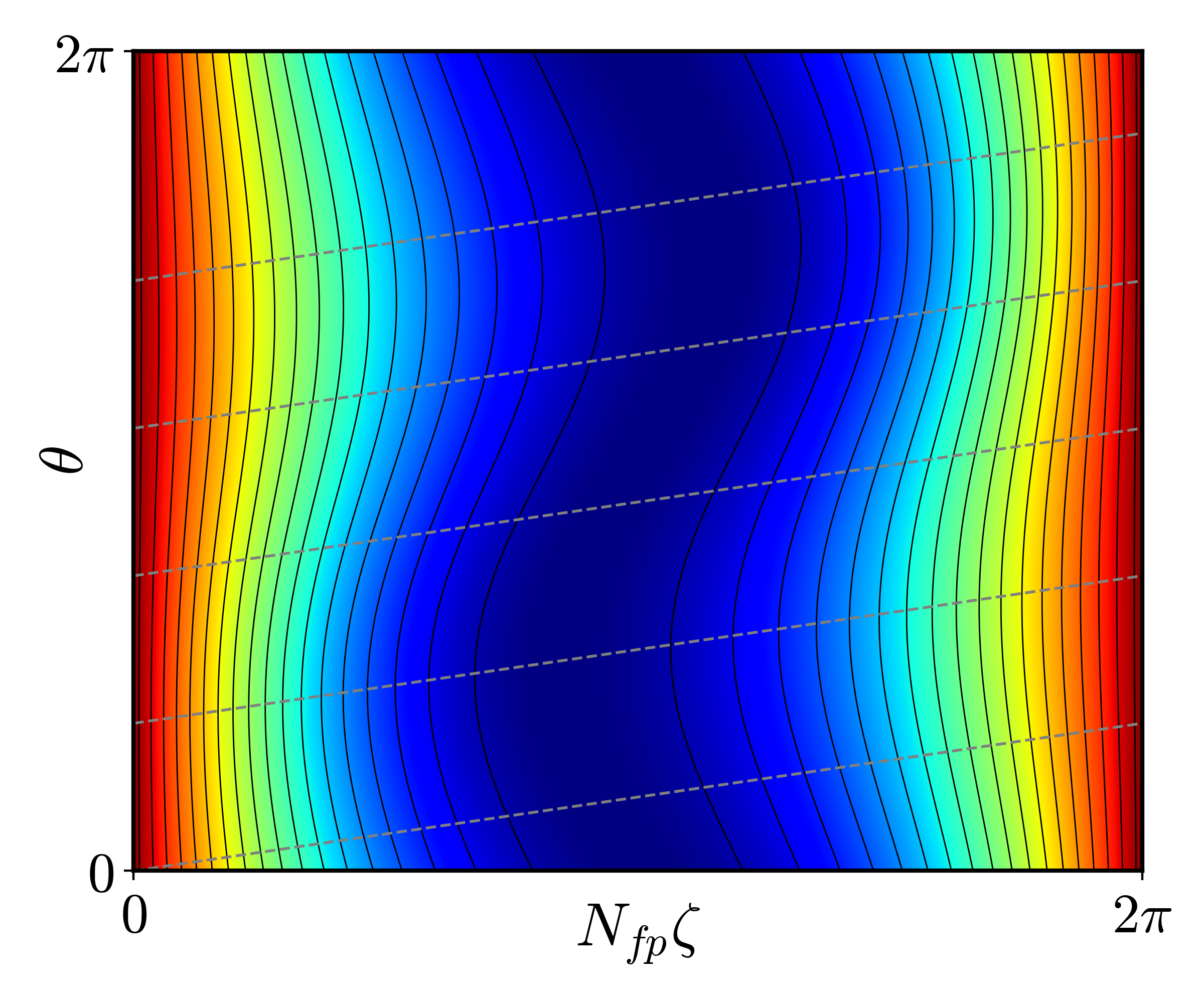}
\includegraphics[angle=0,width=0.49\columnwidth]{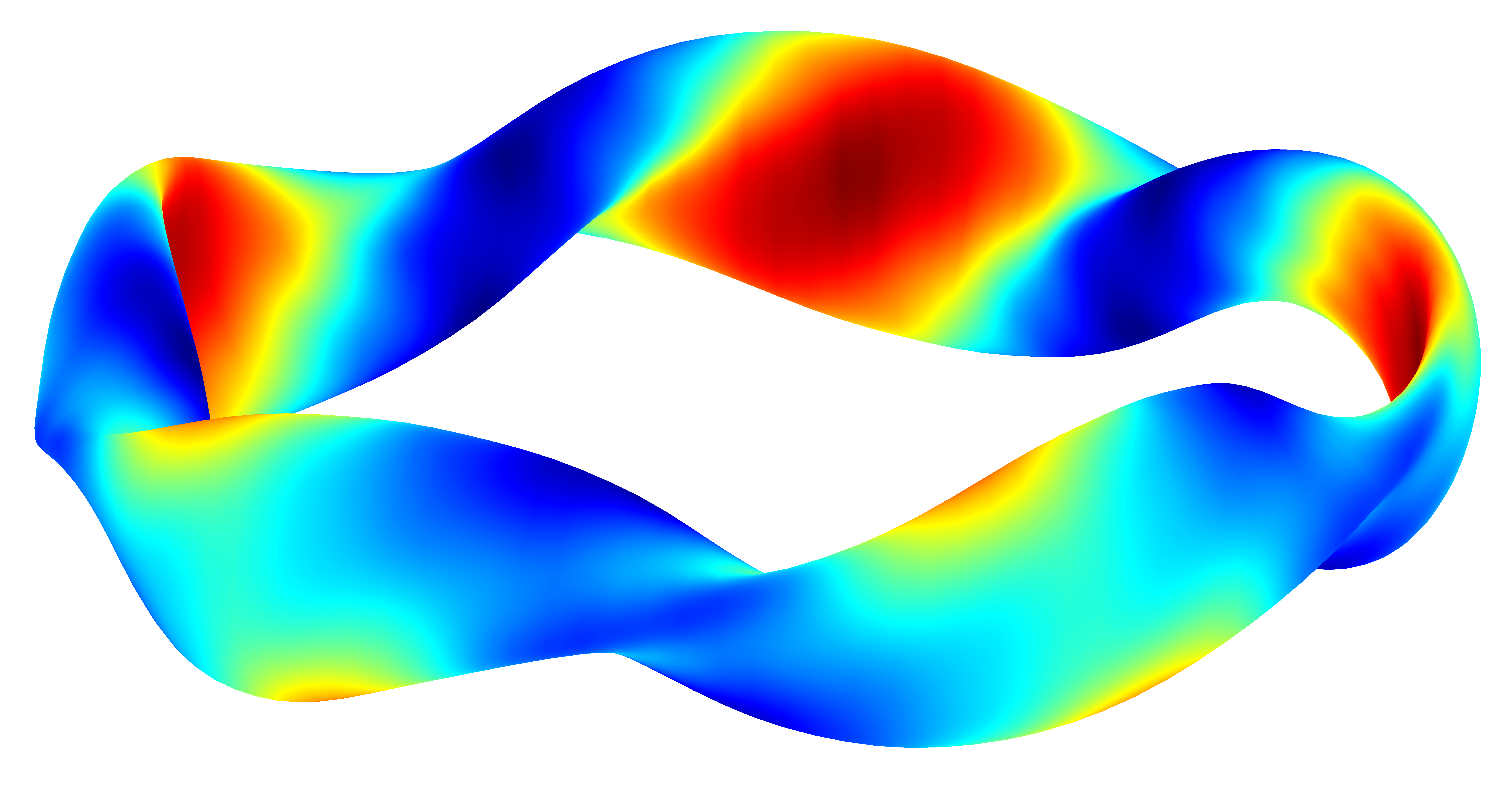}
\caption{$B$ on a flux surface of an axisymmetric tokamak or an exactly quasi-axisymmetric stellarator (top left), of an exactly QS field with helical symmetry (center left), and of an exactly QI field (bottom left); $B$ on the boundary of NCSX (top right), HSX (center right) and the high-mirror configuration of W7-X (bottom right). Dashed gray lines represent magnetic-field lines.\label{FIG_EXAMPLES}}
\end{figure}

Although the stellarator is considered to have been the first magnetic confinement fusion device~\cite{spitzer1958stellarator}, the tokamak has been the leading concept since, in the late sixties, it demonstrated a performance far greater than any other concept~\cite{arcimovich1968tokamak}. This is a consequence of its axial symmetry, which ensures good confinement of collisionless particles, and thus reduced neoclassical transport. Such symmetry does not exist in the stellarator, which produces its magnetic field through a three-dimensional shaping of its magnetic surfaces. While this eliminates the requirement of a large inductive current, and thus facilitates the operation of the stellarator with respect to that of the tokamak, it generally limits its performance~\cite{helander2012stell}. Specifically, low-collisionality neoclassical transport is intolerably large in a generic stellarator.

It was not until the eighties~\cite{boozer1983qs} that it was discovered that, if the flux surface of the stellarator is mapped with the appropriate set of toroidal and poloidal coordinates, $\theta$ and $\zeta$, termed 'Boozer coordinates', the guiding-center trajectories are determined only by the spatial variation of $B$ on the surface. Consequently, if $B$ can be made independent of the toroidal angle as in a tokamak (figure \ref{FIG_EXAMPLES}, top left), these orbits are isomorphic to those in a tokamak, and thus well confined. Such stellarator magnetic fields are termed 'quasi-axisymmetric'. Good confinement is also the case, more generally, if the spatial dependence of $B$ on the flux surface can be written as a function of $M\theta-N\zeta$, being $N$ and $M$ two integer numbers. The first quasisymmetric (QS) magnetohydrodynamic (MHD) equilibria (of the quasi-helically symmetric family, the one with $N$ and $M$ different from zero, figure \ref{FIG_EXAMPLES}, center left) were obtained numerically several years later~\cite{nuhrenberg1988qs}. In the nineties, Cary and Shasharina~\cite{cary1997omni} generalized the concept of quasisymmetry by proposing magnetic fields with collisionless confinement of trapped particles without the requirement that all the $B$-contours have to be straight lines in Boozer coordinates. The key point is that, even in the absence of a symmetry direction, trapped particles moving in a magnetic field still display a conserved quantity, the second adiabatic invariant
\begin{equation}
J \equiv 2\int_{l_{b_1}}^{l_{b_2}}\sqrt{2\left(\mathcal{E}-\mu B\right)}\mathrm{d}l\,.\label{EQ_J}
\end{equation}
Here, $l$ is the arc-length of field lines, and $l_{b_1}$ and $l_{b_2}$ the bounce-point positions. Cary and Shasharina prescribed the two conditions on the variation of $B$ on the flux surface for a magnetic field to be omnigenous. First, all contours of constant $B$ have to be curves that close in the toroidal, poloidal or helical direction (a feature that eliminates transitioning particles\jl{, see figure~\ref{FIG_ORBITS}}). Additionally, the distance along the field line between consecutive bounce points must not depend on the field line for a given flux surface. When this happens, the contours of constant $J$ are aligned with the flux surfaces. Since $J$ is a conserved quantity, this implies that the particles must drift, on average, on the flux-surface. Around that date, approximately quasi-isodynamic (QI) fields (i.e., omnigenous fields with poloidally-closed $B$-contours, figure \ref{FIG_EXAMPLES}, bottom left) were proposed as reactor candidates \cite{wobig1993helias}. QI fields have the additional advantage of displaying zero bootstrap current at low collisionality~\cite{helander2009bootstrap}, which makes them compatible with an island divertor~\cite{sunnpedersen2019divertor} for the management of heat and particle exhaust

\begin{figure}
\includegraphics[angle=0,width=0.95\columnwidth]{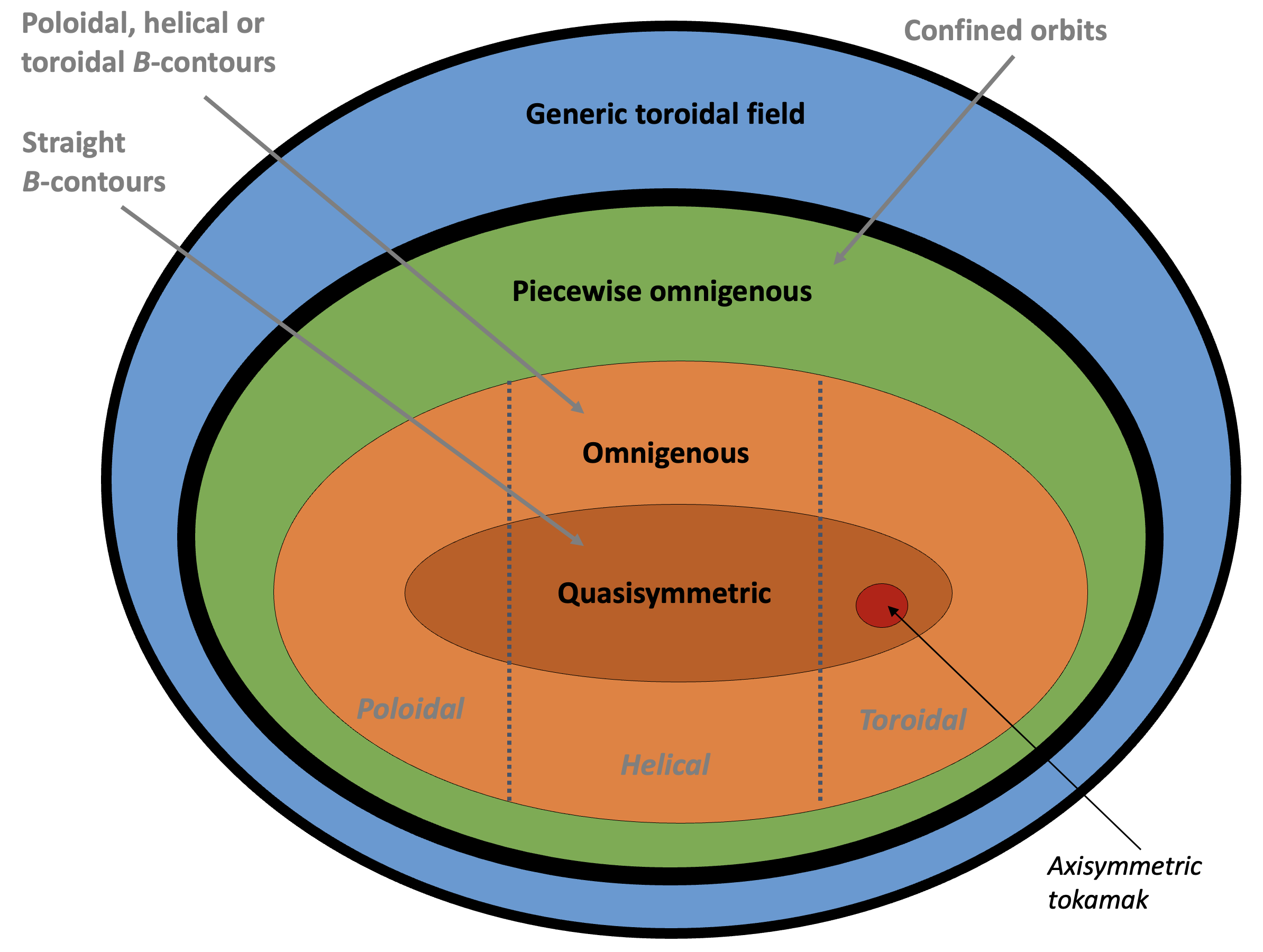}
\includegraphics[angle=0,width=0.49\columnwidth]{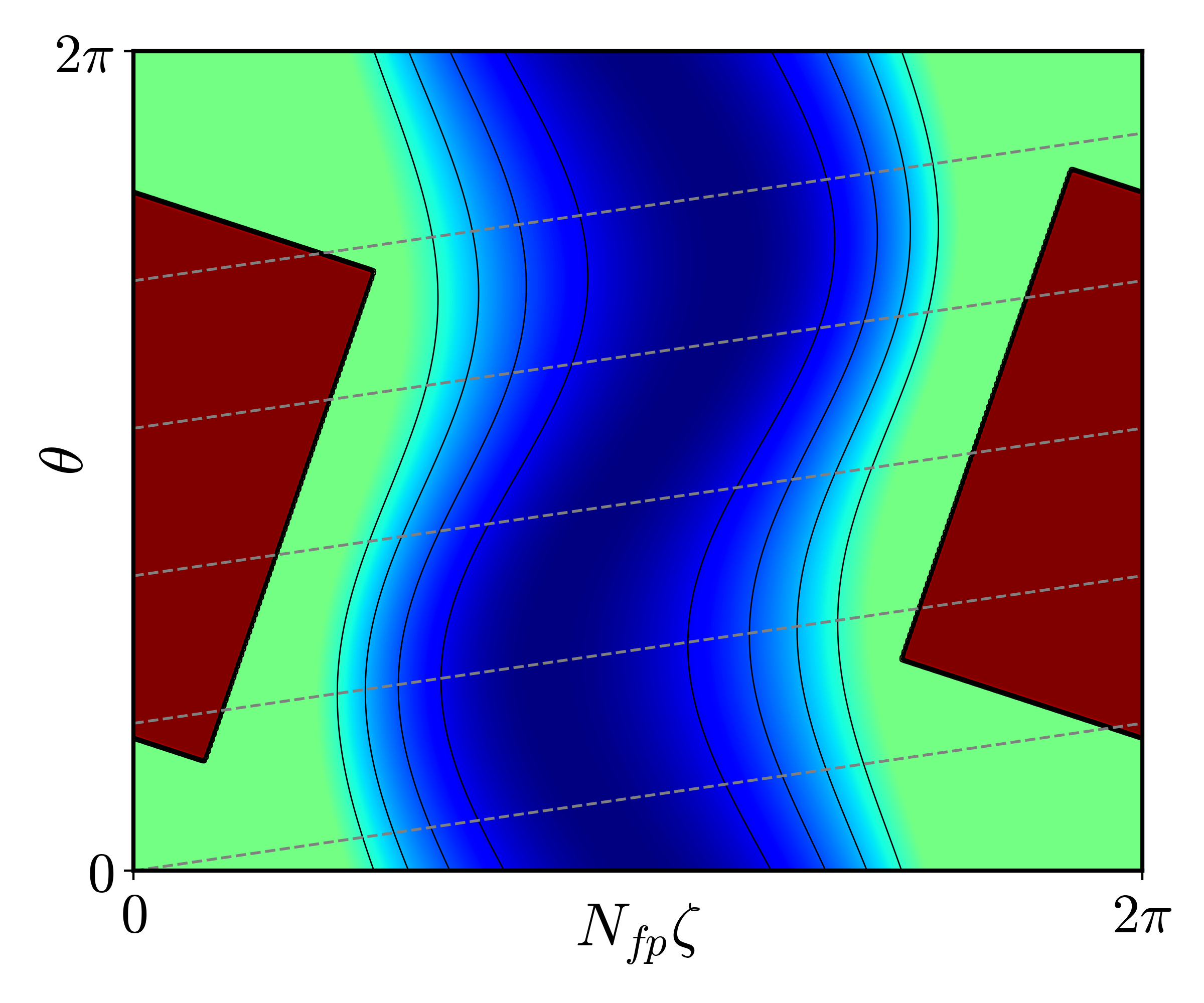}
\includegraphics[angle=0,width=0.49\columnwidth]{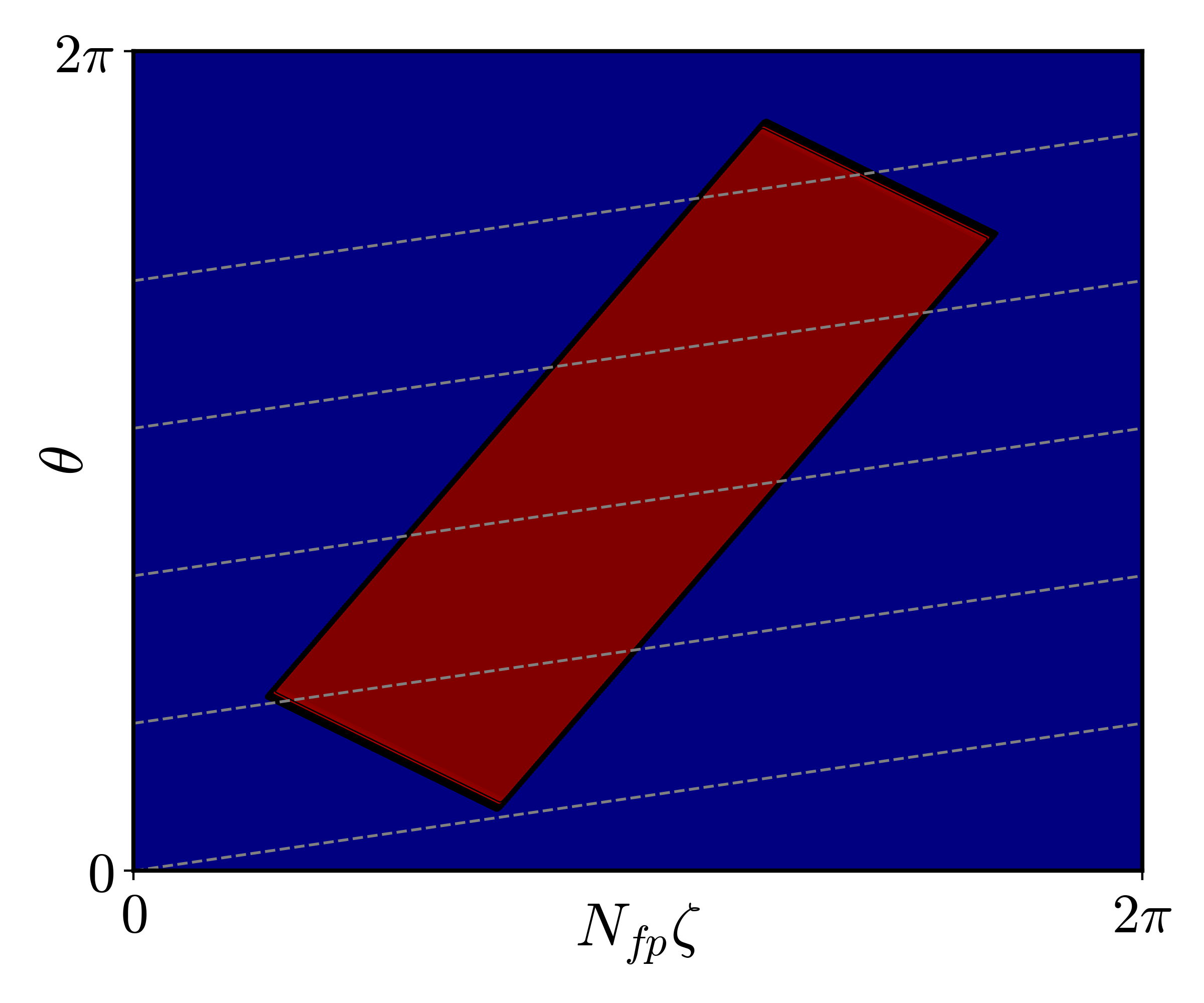}
\caption{Piecewise omnigenity in the space of stellarator magnetic fields: diagram depicting the different families of optimized stellarator fields (top) \jl{and illustrating that quasisymmetric fields are the subset of omnigenous fields with straight $B$-contours, while piecewise omnigenous fields constitute a broader family obtained by relaxing the constraint of toroidally, helically or poloidally-closed $B$-contours}. $B$ on a flux surface of an exactly pwO field close to quasi-isodynamicity (bottom, left) and an exactly pwO field far from omnigenity (bottom, right).\label{FIG_EXAMPLES2}}
\end{figure}

%
%
%

Quasisymmetry and quasi-isodynamicity were promptly employed as design targets, but the first examples of QS and QI configurations presented significant deviations from the ideal limit, and so did the first stellarators ever designed or built according to these criteria, at the beginning of this century: the National Compact Stellarator Experiment (NCSX), the Helically Symmetric eXperiment (HSX) and Wendelstein 7-X (W7-X), see figure \ref{FIG_EXAMPLES} (right). In all cases, violations of the constraints imposed in $B$ by omnigenity and/or quasisymmetry were prominent (note for instance the coil ripple in HSX, or how the contours of high $B$ do not close poloidally in W7-X). As a consequence of this, the neoclassical transport of these devices (particularly that of energetic particles such as fusion-born alpha particles, which are more affected by these imperfections) is too large. Only recently, in this decade, theoretical and computational advances have allowed for configurations with very precise levels of omnigenity~\cite{landreman2022preciseQS,goodman2023qi,dudt2023omni,liu2025omni}. Nevertheless, precise omnigenity should not be expected to be compatible with other reactor relevant criteria. For instance, it is not a guarantee of good energetic particle confinement, see e.g. \cite{goodman2023qi}, as this last property is known to depend critically on the radial variation of $B$~\cite{nemov2008gammac,velasco2021prompt}. Thus, in parallel with these conceptual studies, other, more comprehensive and balanced optimization strategies were being pursued. Finally, almost half a century after the first attempts at stellarator design, this effort has yielded a growing family of candidates for a reactor \cite{sanchez2023qi,goodman2024squids,hegna2025infinity2,lion2025stellaris,swanson2026helios,sanchez2026qi}. Most of them are of the QI type, one of the main reasons being compatibility with an island divertor, see e.g.~\cite{bader2025divertor}

The theoretical development of piecewise omnigenity, on the other hand, is very recent~\cite{velasco2024pwO,velasco2025parapwO,calvo2025pwO}. It generalizes the concept of omnigenity by realizing that the latter can be fulfilled \textit{piecewisely} on the flux surface. Rather than $J$ being a flux-surface constant for a given particle velocity, as in an omnigenous field, the flux surface of a piecewise omnigenous (pwO) field can be divided into several regions, within which $J$ can take different constant values. This eliminates the requirement that all the $B$-contours close in a particular direction \jl{(i.e., the first of the two conditions prescribed by Cary and Shasharina~\cite{cary1997omni}}), thus not necessarily constraining the fields to a limited and discrete set of possible \textit{helicities}, as illustrated by the dashed lines in figure \ref{FIG_EXAMPLES2} (top). Additionally to tokamak-like transport of the bulk species, pwO fields can display zero bootstrap current at low collisionality for arbitrary plasma gradients~\cite{calvo2025pwO}, similarly to QI fields. Two examples of pwO fields are depicted in figure \ref{FIG_EXAMPLES2} (bottom). While figure~\ref{FIG_EXAMPLES2} (bottom, left) represents a pwO field that is relatively close to quasi-isodynamicity (deeply trapped particles encounter $B$-contours that close poloidally), figure~\ref{FIG_EXAMPLES2} (bottom, right) represents a pwO field completely away from omnigenity. \jl{Both cases display wide regions of constant $B$, but it should be noted that smoother examples exist~\cite{velasco2025parapwO}.}

Although piecewise omnigenity is a promising concept, there nevertheless exists a tremendous gap between the level of maturity of pwO fields and that of QI or QS fields in the path towards the fusion reactor. Only a very limited set of configurations have been identified that present a spatial variation of $B$ that is reminiscent of that corresponding to piecewise omnigenity \cite{spong1998jstar,velasco2024pwO,bindel2023direct,liu2025omni,gaur2025umbilical}. Because they are rough, oftentimes coincidental, approximations to piecewise omnigenity, some of them present a too large neoclassical transport (in the absence of the notion of piecewise \jl{omnigenity}, the term quasi-omnigenous \cite{spong1998jstar} was occasionally used, despite lacking a precise definition). Moreover, many of them do not fulfill basic reactor criteria such as MHD stability. In this work, we close this gap by presenting  a stellarator magnetic configuration explicitly designed to satisfy, thanks to unprecedented levels of piecewise omnigenity, the standard set of physics criteria usually required for a viable reactor candidate~\cite{sanchez2023qi,goodman2024squids}.



\section{Optimization with respect to piecewise omnigenity}\label{SEC_OPT}

As illustrated by figure~\ref{FIG_EXAMPLES2} (bottom, left), piecewise omnigenity is a broader family than the fields explored in~\cite{velasco2024pwO} and depicted in figure~\ref{FIG_EXAMPLES2} (bottom, right), and examples exist with a smoother spatial variation of $B$~\cite{velasco2025parapwO}. However, with the goal of assessing the reactor viability of the concept beyond any doubt, in this work we target pwO fields that can be considered as far as possible from omnigenity. As in \cite{velasco2024pwO}, we parametrize the variation of $B$ on the flux surface of a pwO field as:
\begin{eqnarray} 
B_{pwO}(\theta,\zeta)&=&B_{\mathrm{min}}+
(B_{\mathrm{max}}-B_{\mathrm{min}})\times\nonumber\\
& &\hskip-1cm e^{-\left(\frac{\zeta-\zeta_c+t_1(\theta-\theta_c)}{w_1}\right)^{2p}-\left(\frac{\theta-\theta_c+t_2(\zeta-\zeta_c)}{w_2}\right)^{2p}}\,,\label{EQ_BPWO} \\
w_1&=&\frac{\pi}{N_{fp}}\frac{1-t_1t_2}{1+t_2/\iota}\,,\label{EQ_W1}\\
w_2&=&\pi\,.\label{EQ_W2}
\end{eqnarray}
Here, $B_{\mathrm{max}}$ and $B_{\mathrm{min}}$ are the maximum and minimum values of $B$ on the flux surface, $\iota$ is the rotational transform, which determines the direction of the field lines on the flux surface, and $N_{fp}$ is the number of field periods, $B(\theta,\zeta)=B(\theta,\zeta+2\pi/N_{fp})$. In the limit $p\rightarrow\infty$, all the $B$-contours of equation (\ref{EQ_BPWO}) collapse into a parallelogram: $B=B_{\mathrm{max}}$ inside it and $B=B_{\mathrm{min}}$ outside it. The rest of the quantities in  equation~(\ref{EQ_BPWO}) parametrize the shape of this parallelogram. Firstly, $\zeta_c$ and $\theta_c$ determine the position of its center. Then, $w_1$ and $w_2$ specify its toroidal and poloidal extension: $w_1$ is set in equation~(\ref{EQ_W1}) (as a function of other equilibrium quantities, such as $\iota$, that will change during the optimization process) by the requirement of collisionless confinement of trapped particles~\cite{velasco2024pwO}; $w_2$,  see equation~(\ref{EQ_W2}),  is set by the requirement of zero bootstrap current~\cite{calvo2025pwO}. Finally, $t_1$ and $t_2$ determine the slope of the two different sides of the parallelogram ($-1/t_1$ and $-t_2$ respectively).

Equation~(\ref{EQ_W1}) points at a critical dependence of the property of piecewise omnigenity on the value of $\iota$. However, for the family of pwO fields presented in~\cite{velasco2024pwO} and targeted in this work, the $B$-contours are composed of straight lines. For this reason, we do not expect a large degradation of the transport properties to be caused by a moderate change in $\iota$ due e.g. to the bootstrap current. In a field like the one of figure \ref{FIG_EXAMPLES2} (bottom, right), two neighbouring orbits that originally shared a value of $J$ will, very likely, share a (different) value of $J$ after the change of $\iota$, thus keeping $\partial_\alpha J=0$ (here, $\alpha=\theta-\iota\zeta$ is employed to label field lines on a flux surface, and therefore $\partial_\alpha J$ denotes the variation of $J$ within the flux surface). Only the orbits close to the juncture between regions will start to cause transport, if one of the bounce points \textit{jumps} to a different parallelogram segment. Another way to understand this expected robustness against $\iota$ changes is that, within each region of the flux surface, trapped particles behave \textit{as if} they were moving in a QS field (with $N/M$ equal to $1/t_1$ or  $t_2$, depending on the region); the property of zero effective ripple of QS fields, differently to what happens to more general omnigenous fields, does not depend explicitly on $\iota$, see e.g.~\cite{landreman2012omni}.

The MHD equilibrium of a stellarator is determined in fixed-boundary simulations by the plasma pressure profile (in this work, the normalized toroidal flux, $s=\psi/\psi_\mathrm{boundary}$, is employed as the radial coordinate, and the pressure profile is set to be linear, with normalized volume-averaged pressure $\beta=0.5\%$), by the toroidal current through the plasma (which we set to zero), and by the shape of the equilibrium boundary. Stellarator optimization is typically carried out by continuously modifying the shape of the boundary surface in order to explore (a region of) the stellarator configuration space. The optimizer performs such exploration by trying to minimize a cost function in which a set of desired properties have been encoded. Ideally, if a minimum of the cost function is found, the corresponding stellarator equilibrium is close to having the desired properties. 

In this work, we have developed specific routines in the code \texttt{DESC} \cite{dudt2023omni} devoted to minimizing the distance of the configuration magnetic field strength, at a specific flux surface $s=s_0$, to the closest field described by equations~(\ref{EQ_BPWO}),~(\ref{EQ_W1}) and~(\ref{EQ_W2}). Specifically, the quantity
\begin{eqnarray} 
\delta B(\theta,\zeta)=\frac{|B(s=s_0,\zeta,\theta)-B_{pwO}(\theta,\zeta)|}{B_{pwO}(\theta,\zeta)}\label{EQ_ERRB}
\end{eqnarray}
is targeted to be zero. During the optimization, in addition to the shape of the $s=1$ flux surface, some of the parameters of equation (\ref{EQ_BPWO}) are varied. We fix $N_{fp}=5$, $\zeta_c=\pi/N_p$ and $\theta_c=0$ (a natural choice given the initial condition, see Appendix~\ref{SEC_DESC}), $s_0=0.5$ and $p=2$. The latter choice implies that, rather than aiming for a precise pwO field, we will actually be aiming for an approximately pwO field that we expect to have (respective to the $p\rightarrow\infty$ case) a smoother spatial variation and a \jl{non-zero but} sufficiently low level of neoclassical transport \jl{(both energy flux and bootstrap current)} for reactor-relevant collisionalities. The other parameters, $B_{\mathrm{max}}$, $B_{\mathrm{min}}$, $t_1$ and $t_2$ (and $\iota$), are left to vary in order not to unnecessarily overconstrain the optimization problem.

\begin{figure}[h!]
\includegraphics[angle=0,width=\columnwidth]{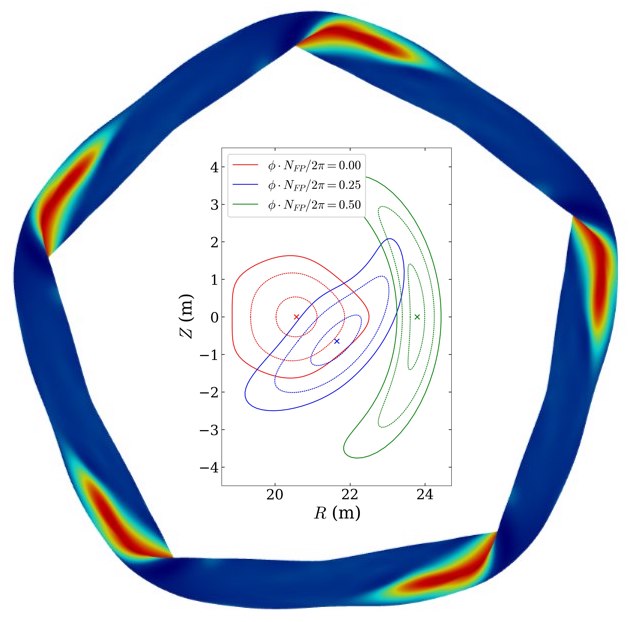}
\includegraphics[angle=0,width=\columnwidth]{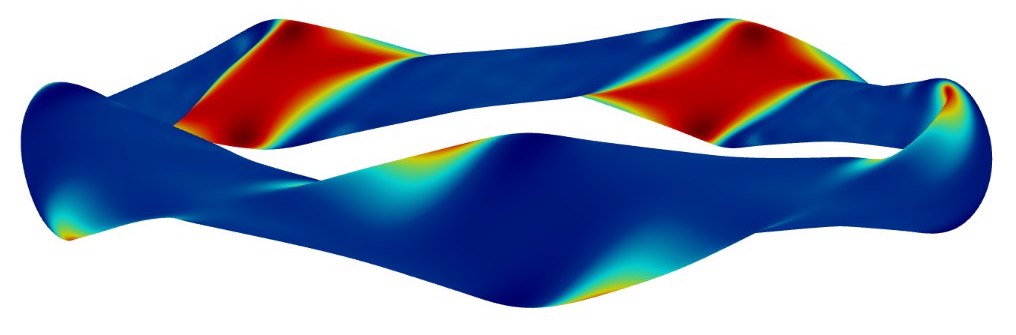}
\caption{Top and side views (top and bottom figures, respectively) of the magnetic field strength on the $s=1$ flux surface of CIEMAT-pw1 at $\beta=0.5\%$, with the shape of several toroidal cuts of the flux surfaces ($R$, $Z$ and $\phi$ are cylindrical coordinates).\label{FIG_EQ3D}}
\end{figure}

\begin{figure*}
\includegraphics[angle=0,width=0.49\columnwidth]{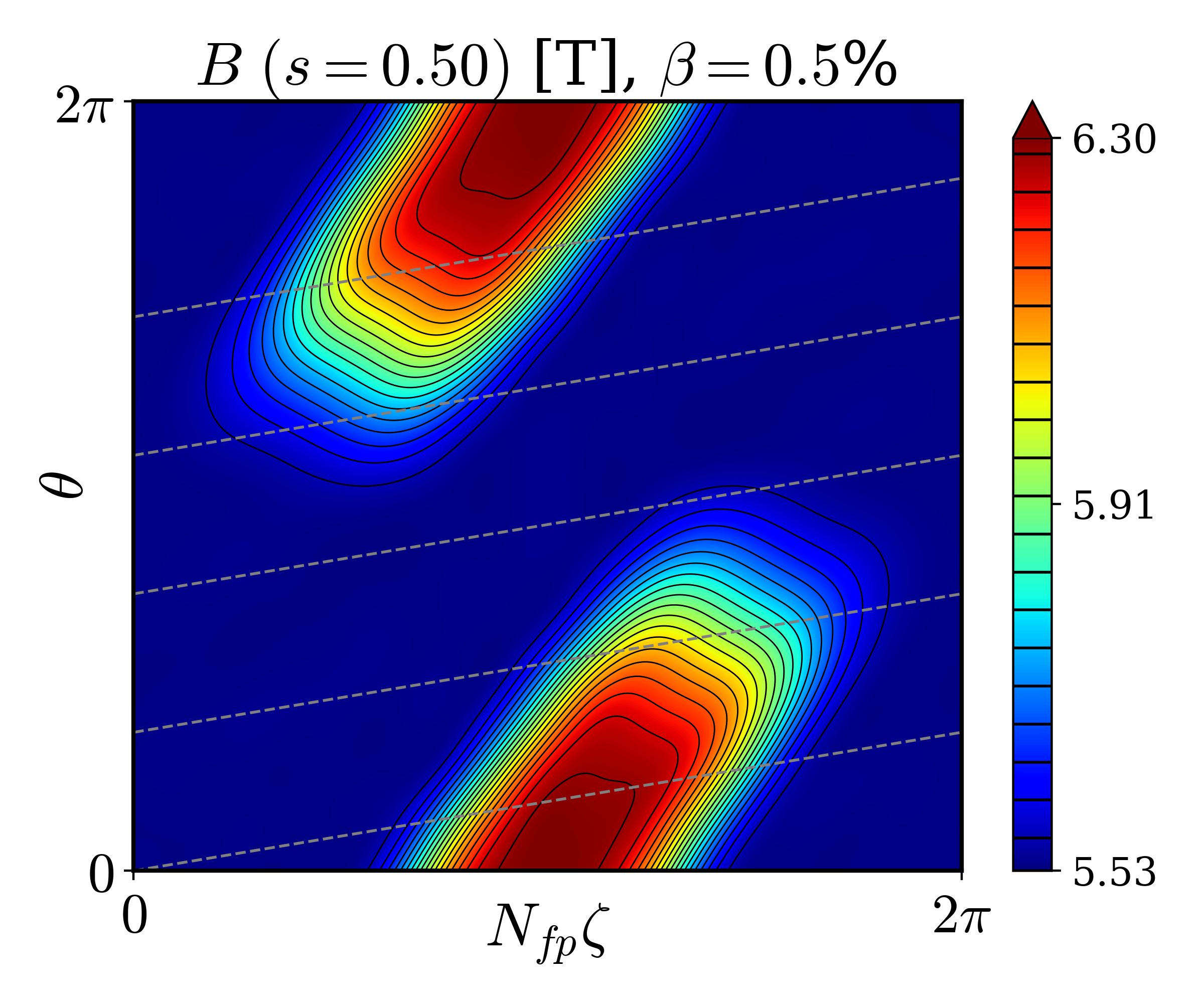}
\includegraphics[angle=0,width=0.49\columnwidth]{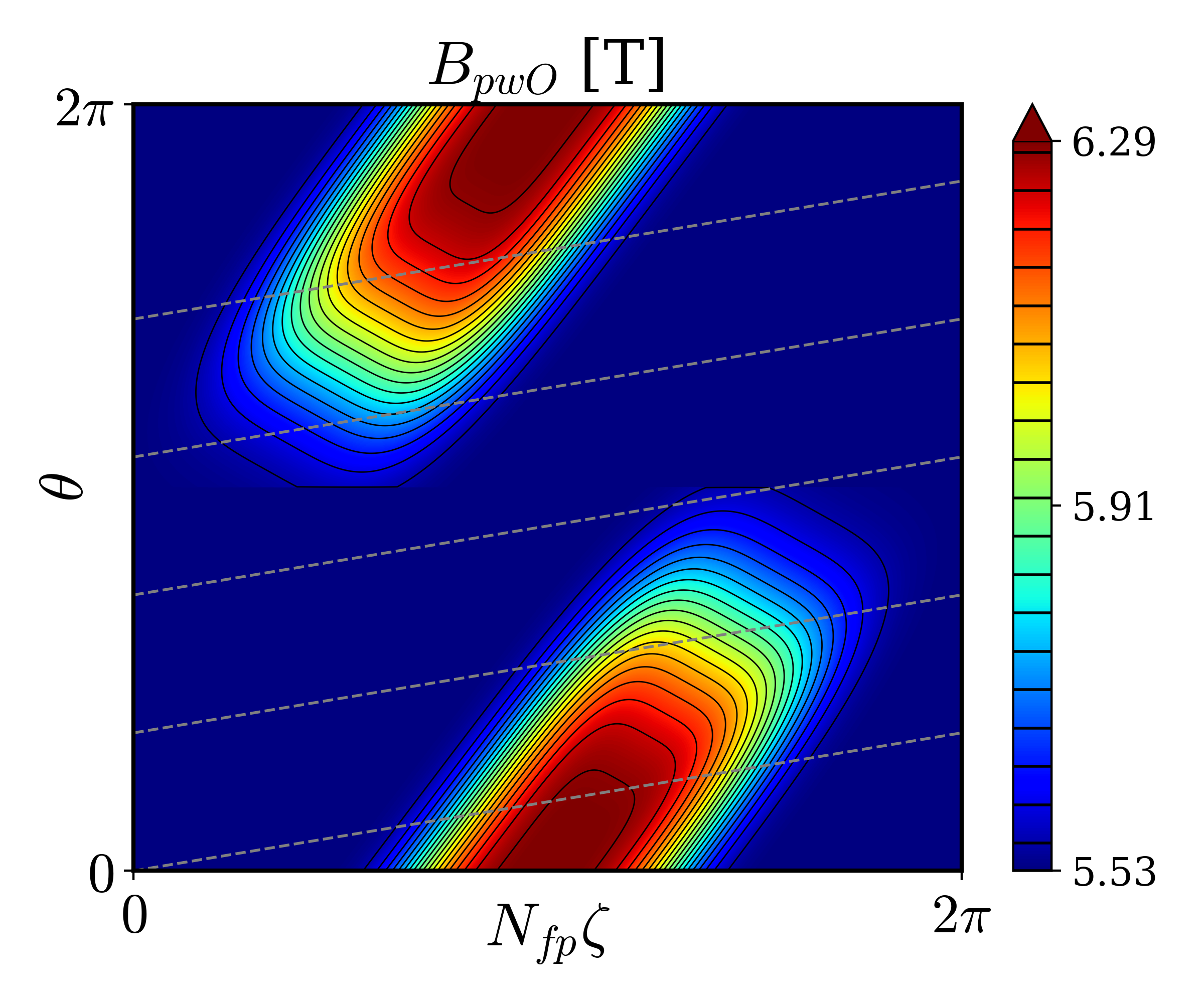}
\includegraphics[angle=0,width=0.49\columnwidth]{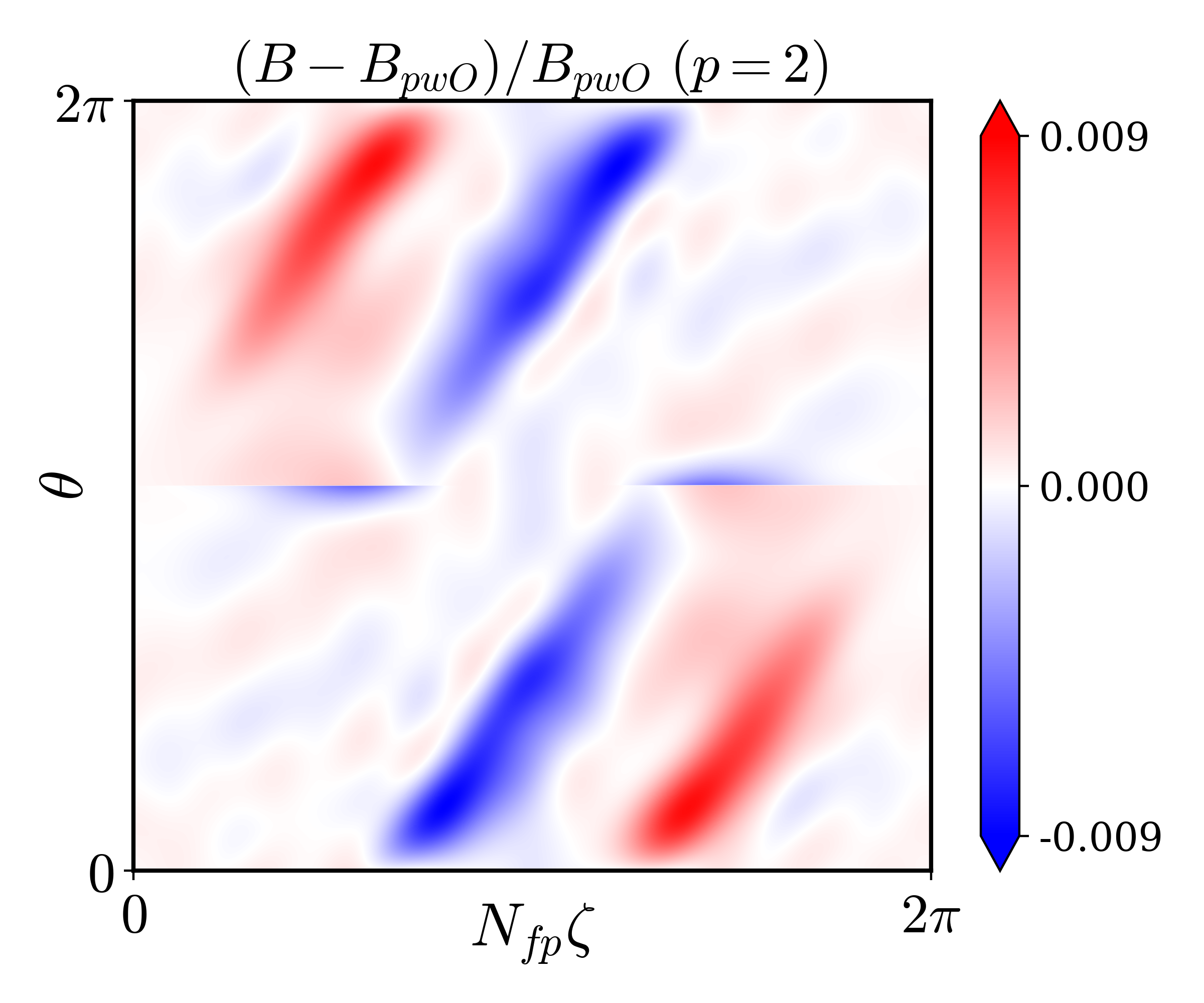}
\includegraphics[angle=0,width=0.49\columnwidth]{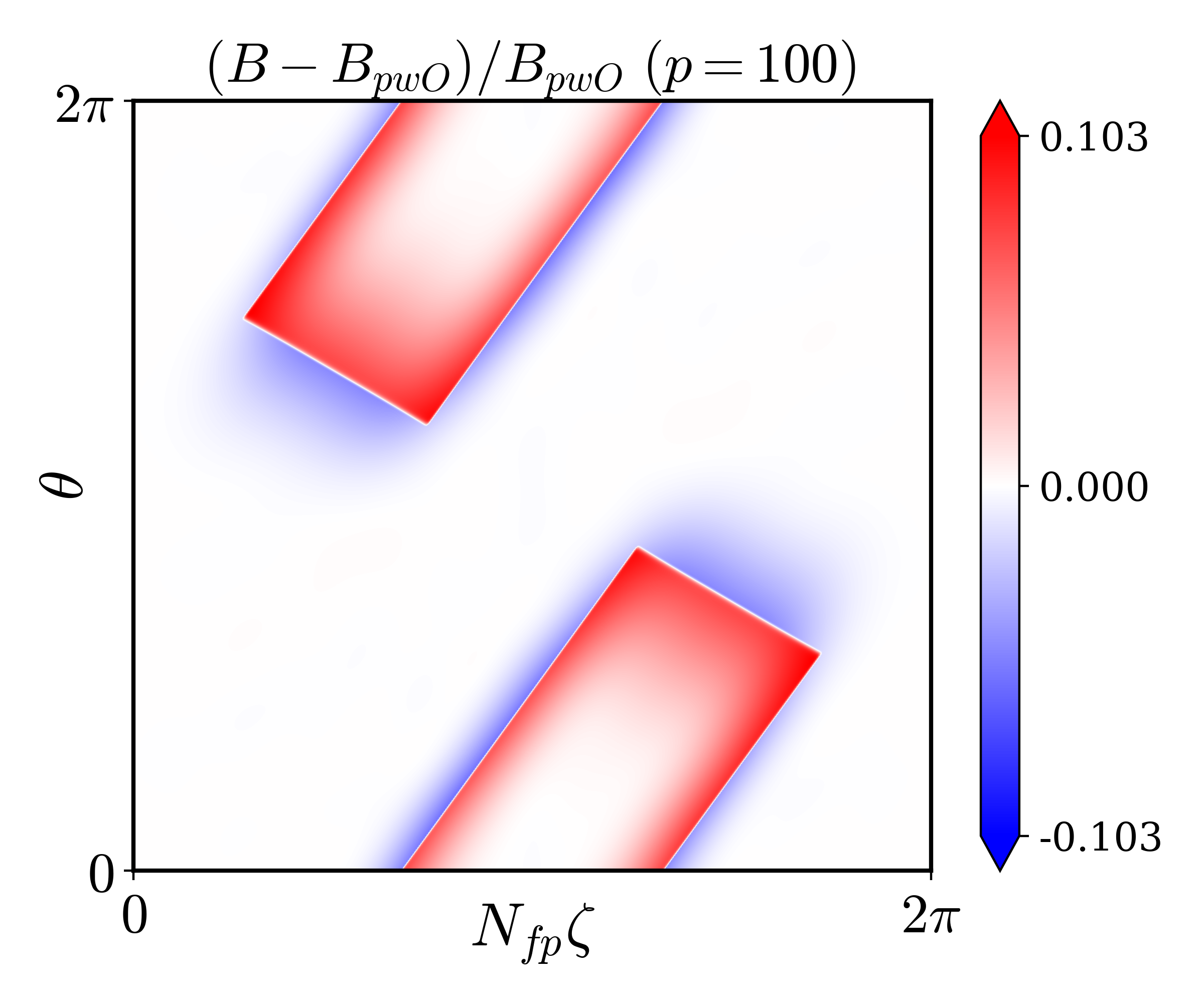}
\includegraphics[angle=0,width=0.49\columnwidth]{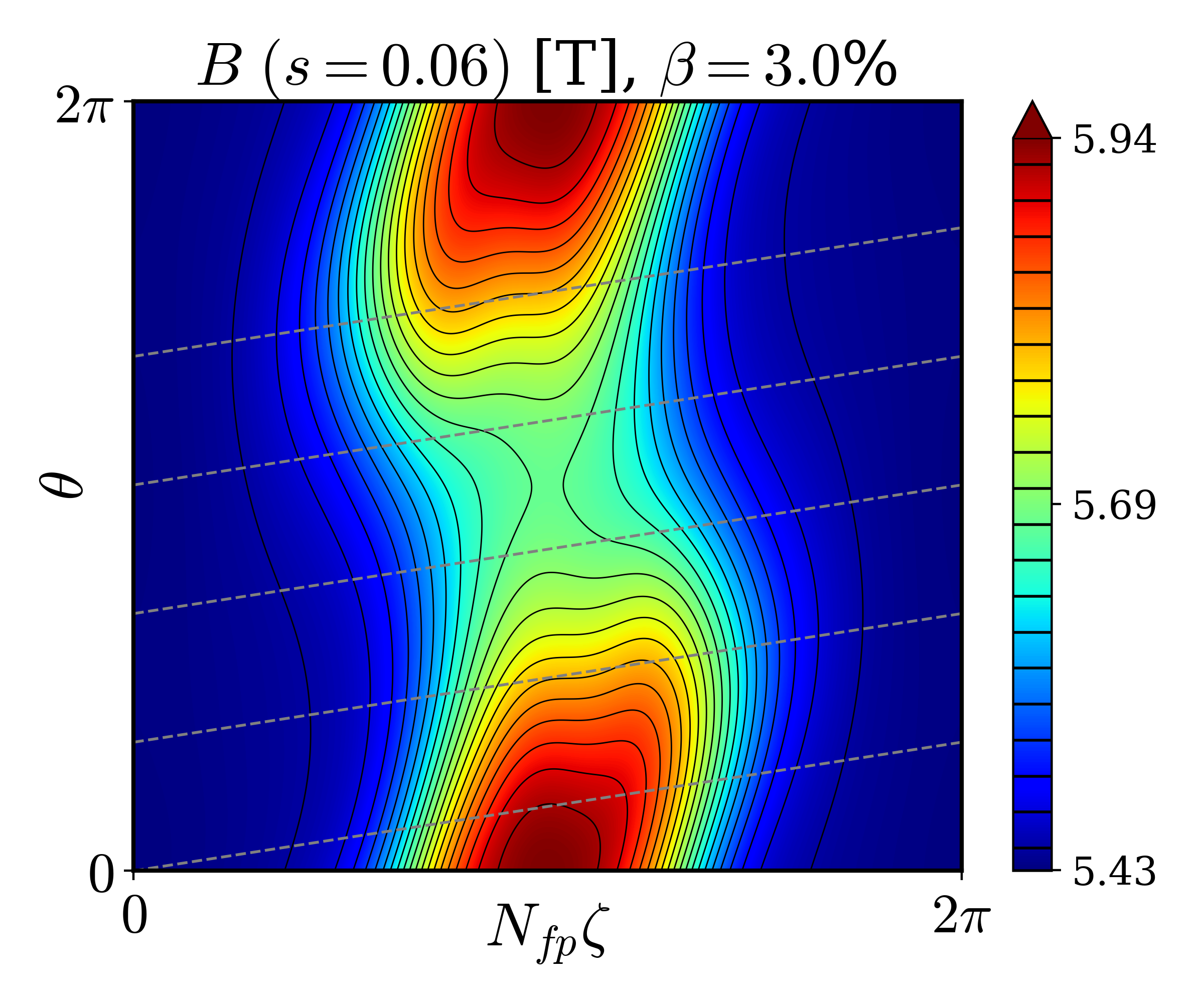}
\includegraphics[angle=0,width=0.49\columnwidth]{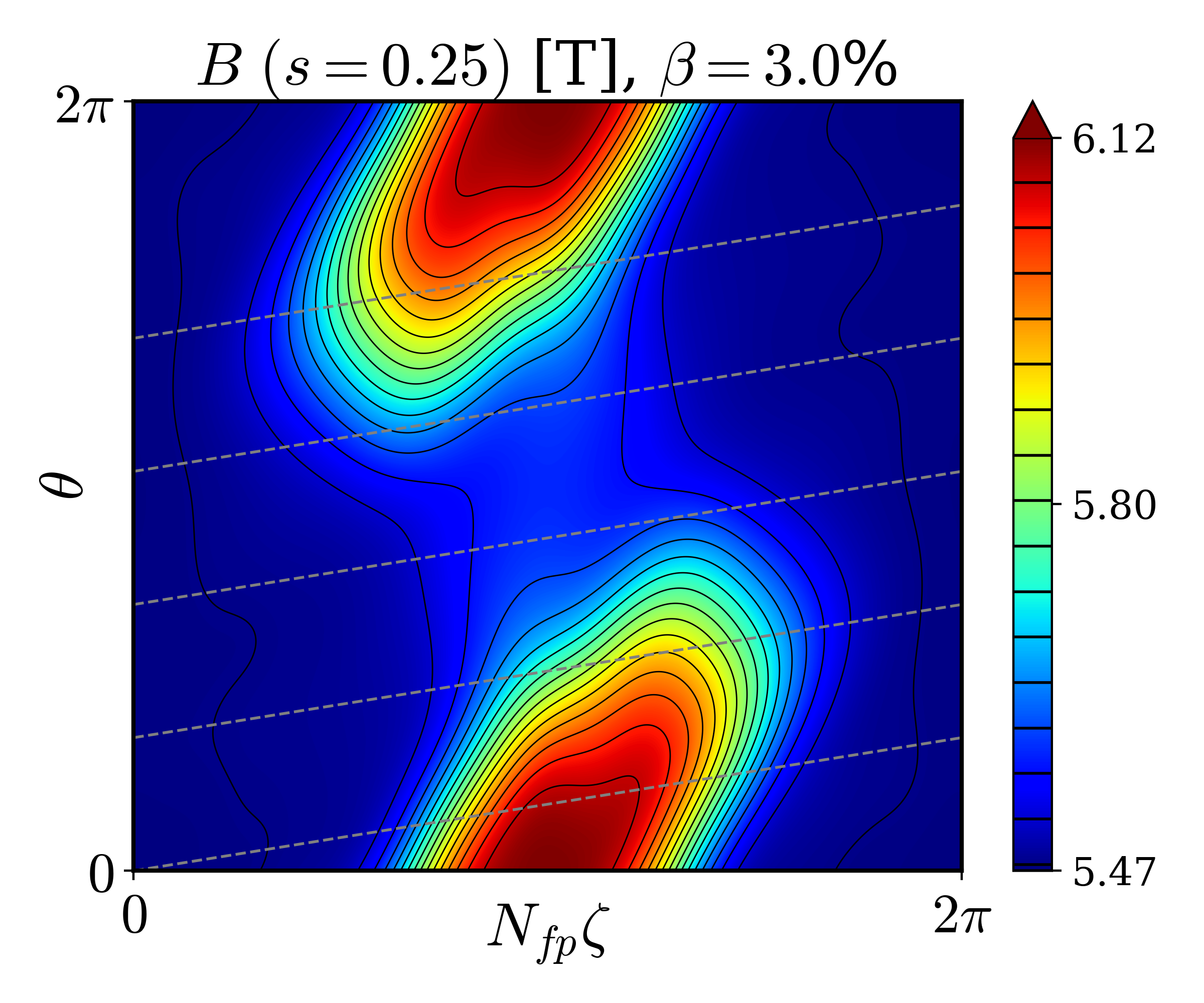}
\includegraphics[angle=0,width=0.49\columnwidth]{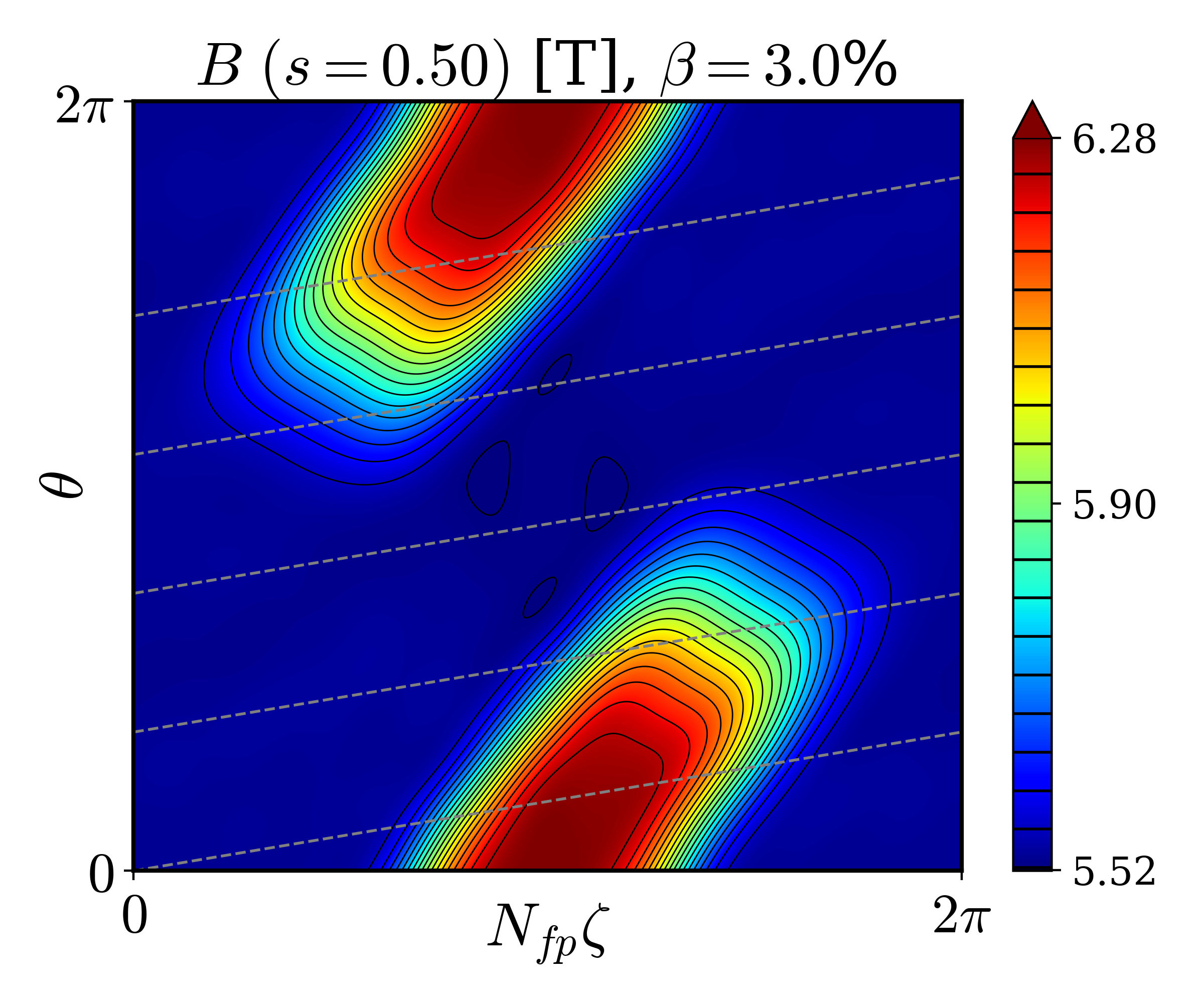}
\includegraphics[angle=0,width=0.49\columnwidth]{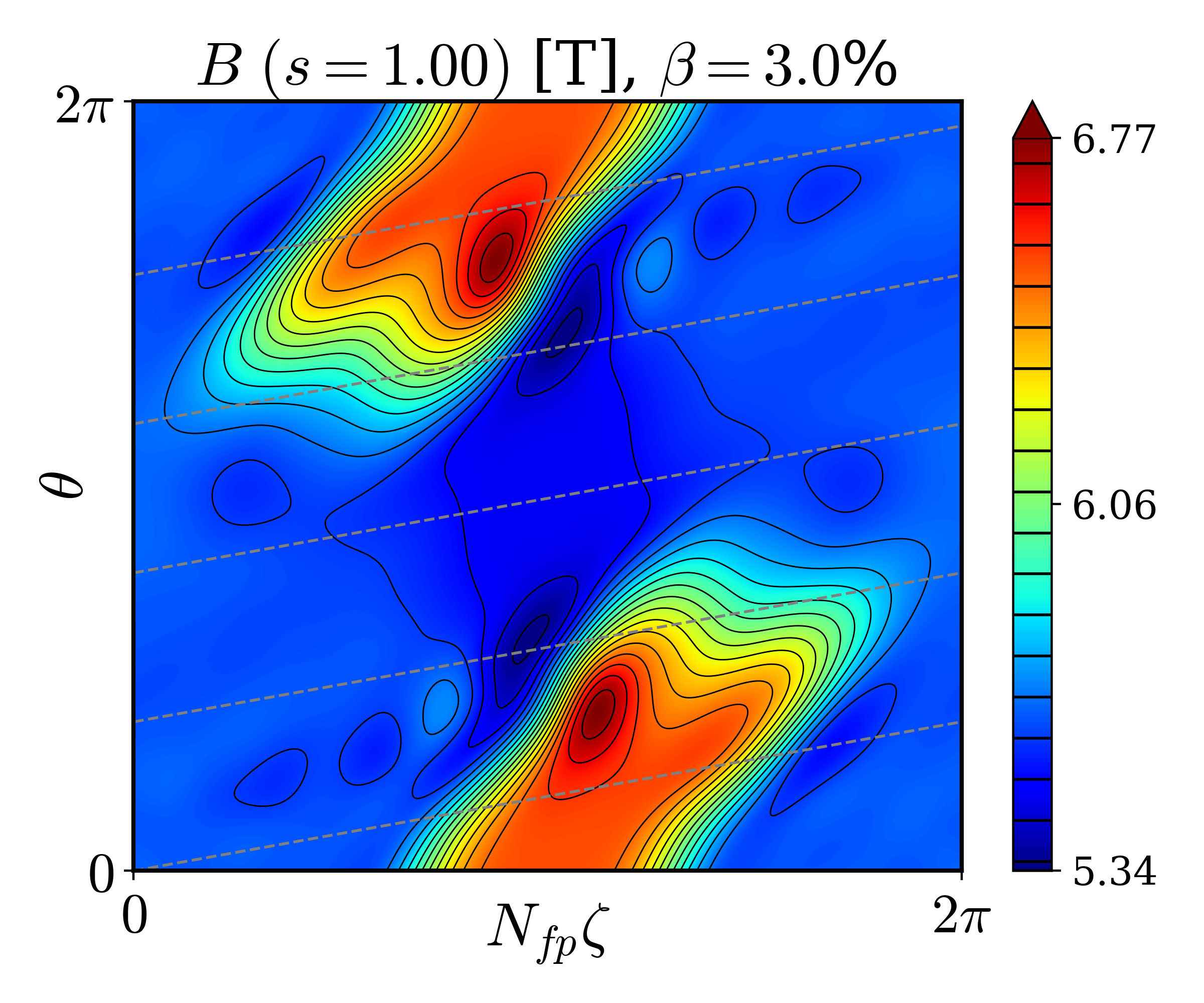}
\caption{Closeness of CIEMAT-pw1 to piecewise omnigenity: Top, from left to right: $B(\theta,\zeta)$ for $\beta=0.5$\% at the radial position $s=0.5$ of the optimized equilibrium; $B_{pwO}(\theta,\zeta)$ at the last iteration of the optimization; relative difference between them, and $p\rightarrow\infty$ limit of the relative difference. Bottom, from left to right:  $B(\theta,\zeta)$ for $\beta=3.0$\% at radial positions $s=0.06$, $s=0.25$, $s=0.50$ and $s=1.00$.\label{FIG_BPWO}}
\end{figure*}

In this work, the quantity $\delta B(\theta,\zeta)$ is incorporated into a cost function together with other reactor-relevant physics properties of the MHD equilibrium. Ideally, the result of the optimization process, once the optimizer has minimized the cost function, should be a magnetic configuration that is close to piecewise omnigenity at $s=s_0$ (with the consequent reduction of neoclassical transport) and, at the same time, satisfies the rest of desired reactor-relevant physics criteria. Achievement of small $\delta B$ will be demonstrated in section~\ref{SEC_BPWO}, and the fulfillment of relevant reactor design criteria, such as an appropriate $\iota$ profile, will be the focus of section~\ref{SEC_REACTOR}. 

There are, however, some physics criteria whose achievement may be facilitated (although not guaranteed) by optimization with respect to small  $\delta B$, and they will appear in both sections. After all, properties such as MHD stability, turbulence and fast ion confinement can be straightforwardly connected, to some extent, to the spatial variation of $B$, both on the flux surface and in the radial direction (in particular, as in the case of omnigenity, good fast ion confinement is significantly facilitated but not guaranteed by closeness to piecewise omnigenity). In the rest of this section, we discuss to what extent optimization with respect to piecewise omnigenity can benefit these aspects of reactor performance.

Firstly, it should be noted that, in this work, piecewise omnigenity has been targeted at a single flux surface, $s_0=0.5$. Similarly to what happens in omnigenous fields, closeness to piecewise omnigenity is expected to degrade as one separates from \jl{the optimized} flux surface \jl{(this advises against trying to make $\delta B(\theta,\zeta)$ very small on two different flux surfaces)}. In particular, the Fourier spectrum of $B_{pwO}$ is quite broad, and $B_{mn}$ modes with different $m$ number are expected to have a different radial variation since, at least sufficiently close to the axis, $B_{mn}\sim s^{m/2}$ (here, $B_{pwO}(\theta,\zeta) = \sum_{m,n} B_{m,n}\cos(m\theta -nN_{p}\zeta)$). It will be shown in section~\ref{SEC_BPWO}  that this degradation is weak enough so that the benefits obtained from optimizing at $s=0.5$ extend to the full plasma volume.

Let us then assume that the magnetic field of a nearly pwO configuration can be approximately written as
\begin{eqnarray} 
B_{pwO}(s,\theta,\zeta)&=&B_{\mathrm{min}}(s)+
(B_{\mathrm{max}}(s)-B_{\mathrm{min}}(s))\times\nonumber\\
& &\hskip-2cm \lim_{p\rightarrow\infty} e^{-\left(\frac{\zeta-\zeta_c+t_1(\theta-\theta_c)}{w_1}\right)^{2p}-\left(\frac{\theta-\theta_c+t_2(\zeta-\zeta_c)}{w_2}\right)^{2p}}\,,\label{EQ_BSTZ}
\end{eqnarray}
with $w_1$ and $w_2$ set by equations (\ref{EQ_W1}) and (\ref{EQ_W2}) and all the parameters in equation (\ref{EQ_BSTZ}) (and $\iota$) kept constant throughout the plasma volume except $B_\mathrm{max}$ and $B_\mathrm{min}$. Then, if $\partial_s B_\mathrm{min}$ and $\partial_s B_\mathrm{max}$ have the same sign and are both sufficiently large (in absolute value), a enhanced trapped-particle precession on the flux surface, at a stronger rate than in the radial direction, is expected to contribute to good alpha-particle confinement~\cite{nemov2008gammac,velasco2021prompt}. Moreover, if  $\partial_s B_\mathrm{max}$ and $\partial_s B_\mathrm{min}$ are both positive, trapped electrons precess in the direction opposite to the electron diamagnetic direction; this mitigates turbulence originated by trapped-electron modes driven by density gradients, and other ion-scale gyrokinetic instabilities involving kinetic electrons~\cite{rosenbluth1968maxJ,helander2013tem,plunk2017maxJ,proll2022maxJ}. This can be understood in terms of the second adiabatic invariant: straightforward calculations analogous to those of section 3.5 of~\cite{velasco2025parapwO} show that, for the magnetic field of equation (\ref{EQ_BSTZ}), if $\partial_s B_\mathrm{min}$ and $\partial_s B_\mathrm{max}$ are positive,
\begin{equation}
J= J^\mathrm{(w)}(s,{\mathcal{E}},\mu),\quad \partial_s J^\mathrm{(w)}<0,\quad\mathrm{w}= \mathrm{I,II,III,...}\label{EQ_MAXJpwO}\\
\end{equation}
\jl{where $\mathrm{w}$ labels the different regions of the flux surface within which $J$ is constant.} That is, the so-called maximum-$J$ property~\cite{rosenbluth1968maxJ,helander2013tem} is fulfilled \textit{piecewisely}. Through the connection between orbit-averated drifts and the spatial variation of $J$, it is straightforward to see that equation~(\ref{EQ_MAXJpwO}) guarantees that both trapped ions and trapped electrons are collisionlessly confined, and that the latter precess in the direction that mitigates said turbulence. In this work we try to satisfy equation (\ref{EQ_MAXJpwO}) by minimizing $\delta B(\theta,\zeta)$ and simultaneously \jl{achieving $\partial_s B_{\mathrm{max}}>0$ and $\partial_s B_{\mathrm{min}}>0$. As we will see in section~\ref{SEC_BPWO}, the latter requirement is met by approximately pwO fields at large $\beta$ thanks to the diamagnetic effect, without the need for a dedicated optimization target. In section~\ref{SEC_REACTOR}, the beneficial effect of this approach towards the maximum-$J$ property will be tested with gyrokinetic simulations.}

Finally, we note that the maximum-$J$ property has been connected (through the effect of the magnetic well) to MHD stability, see e.g.~\cite{rodriguez2024maxJ}.

\begin{figure}
\includegraphics[angle=0,width=0.95\columnwidth]{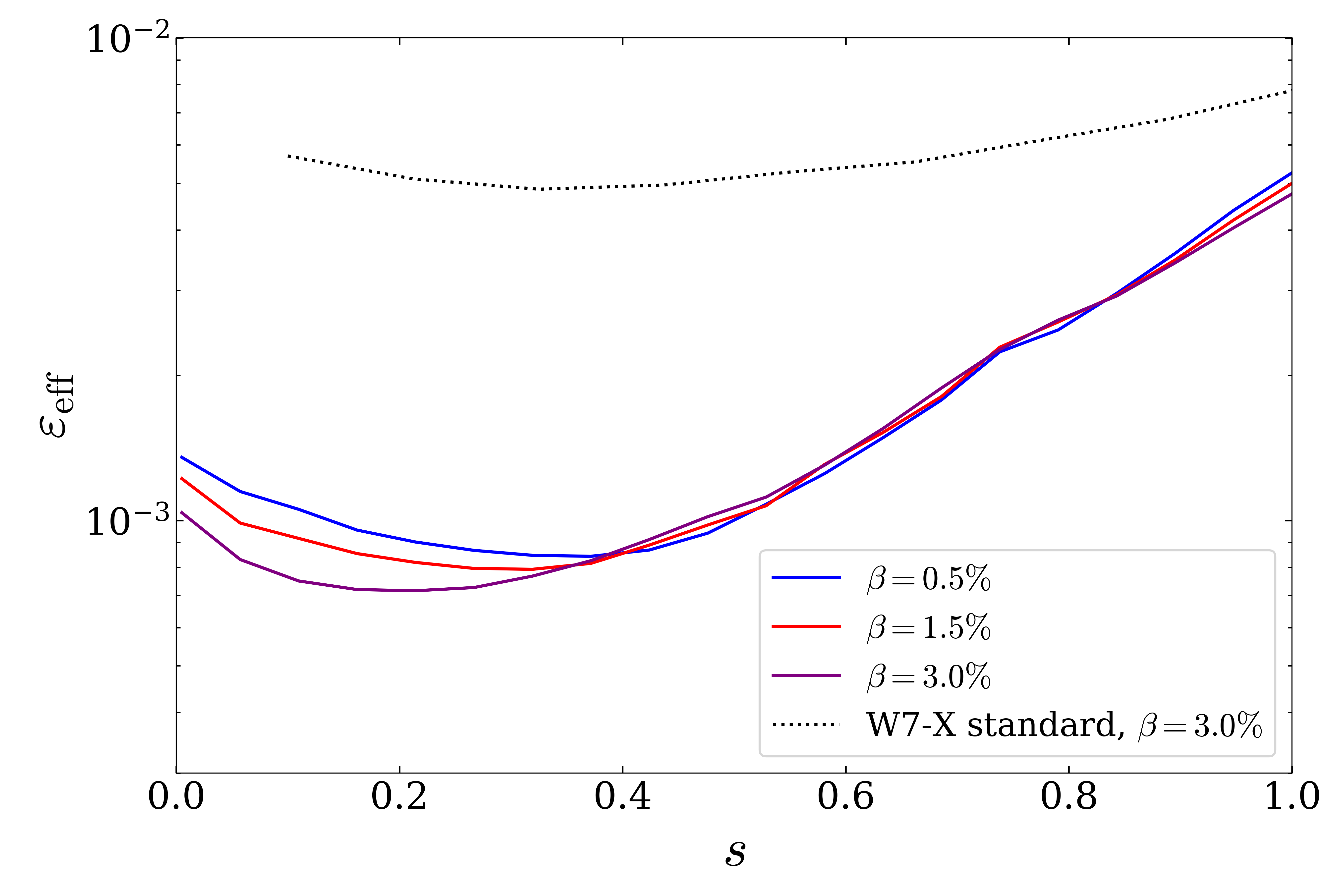}
\includegraphics[angle=0,width=0.95\columnwidth]{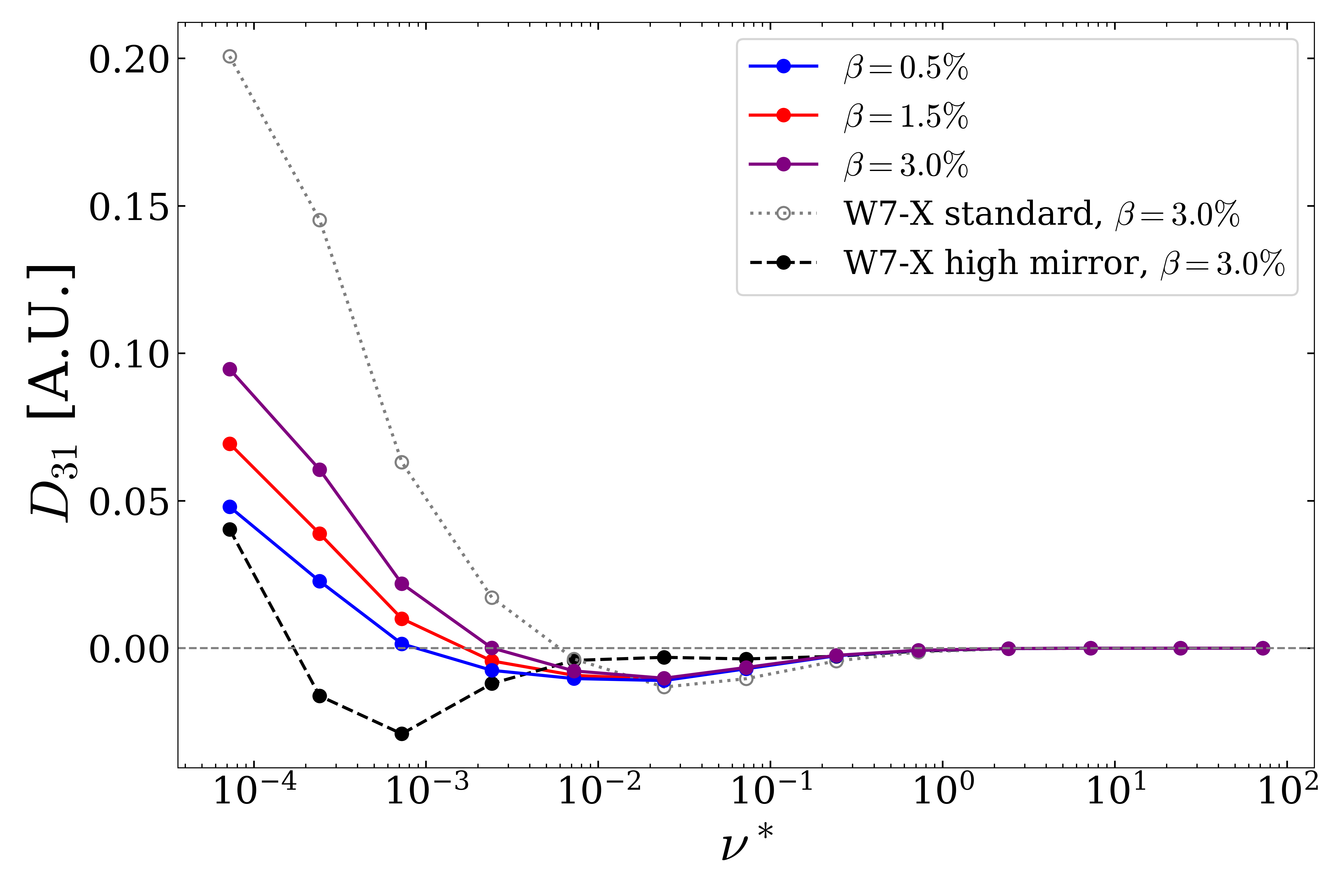}
\caption{Neoclassical transport of CIEMAT-pw1: radial profile of the effective ripple (top) and collisionality dependence of the bootstrap transport coefficient at $s=0.06$ (bottom).\label{FIG_NC}}
\end{figure}

\begin{figure}
\includegraphics[angle=0,width=0.95\columnwidth]{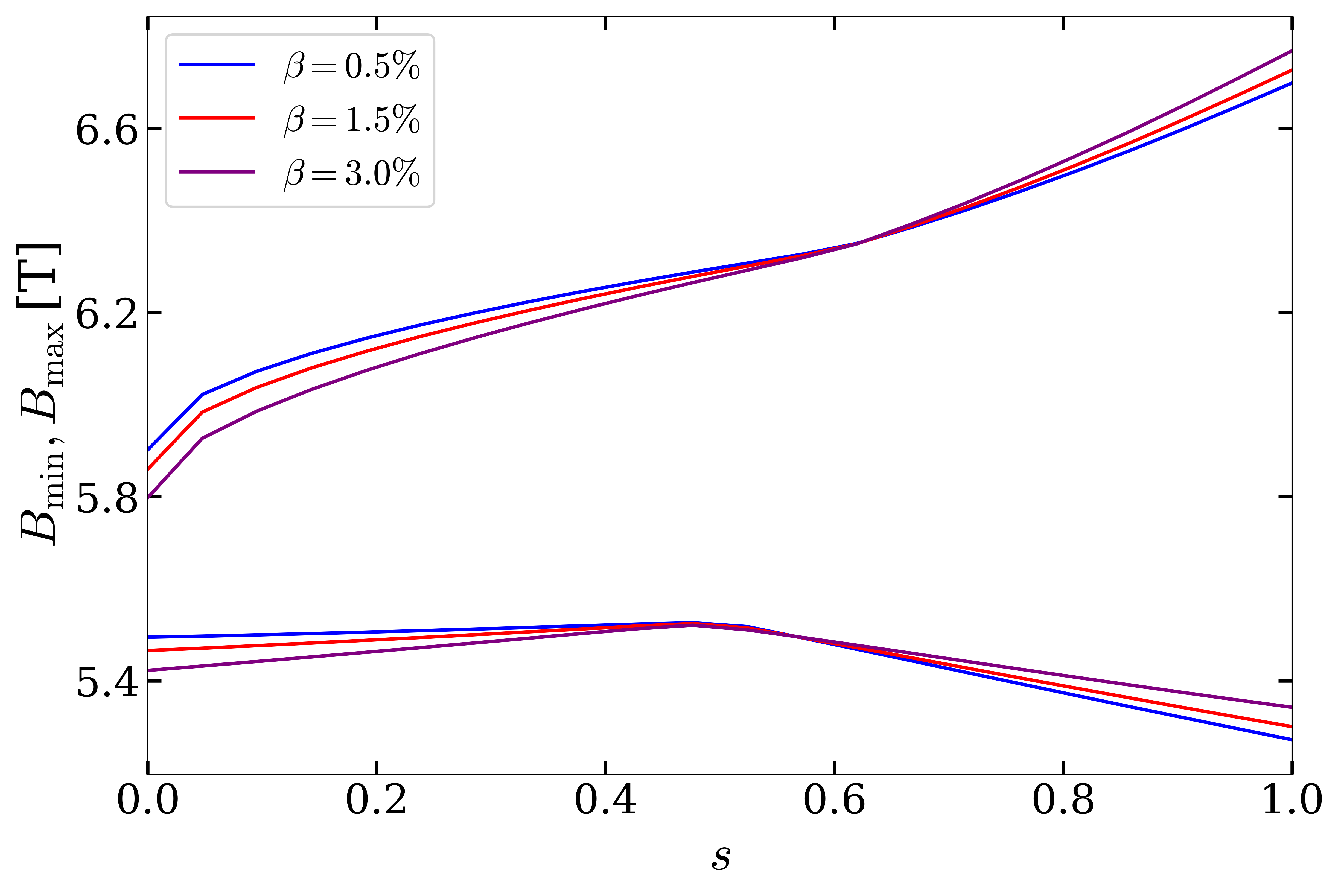}
\includegraphics[angle=0,width=0.95\columnwidth]{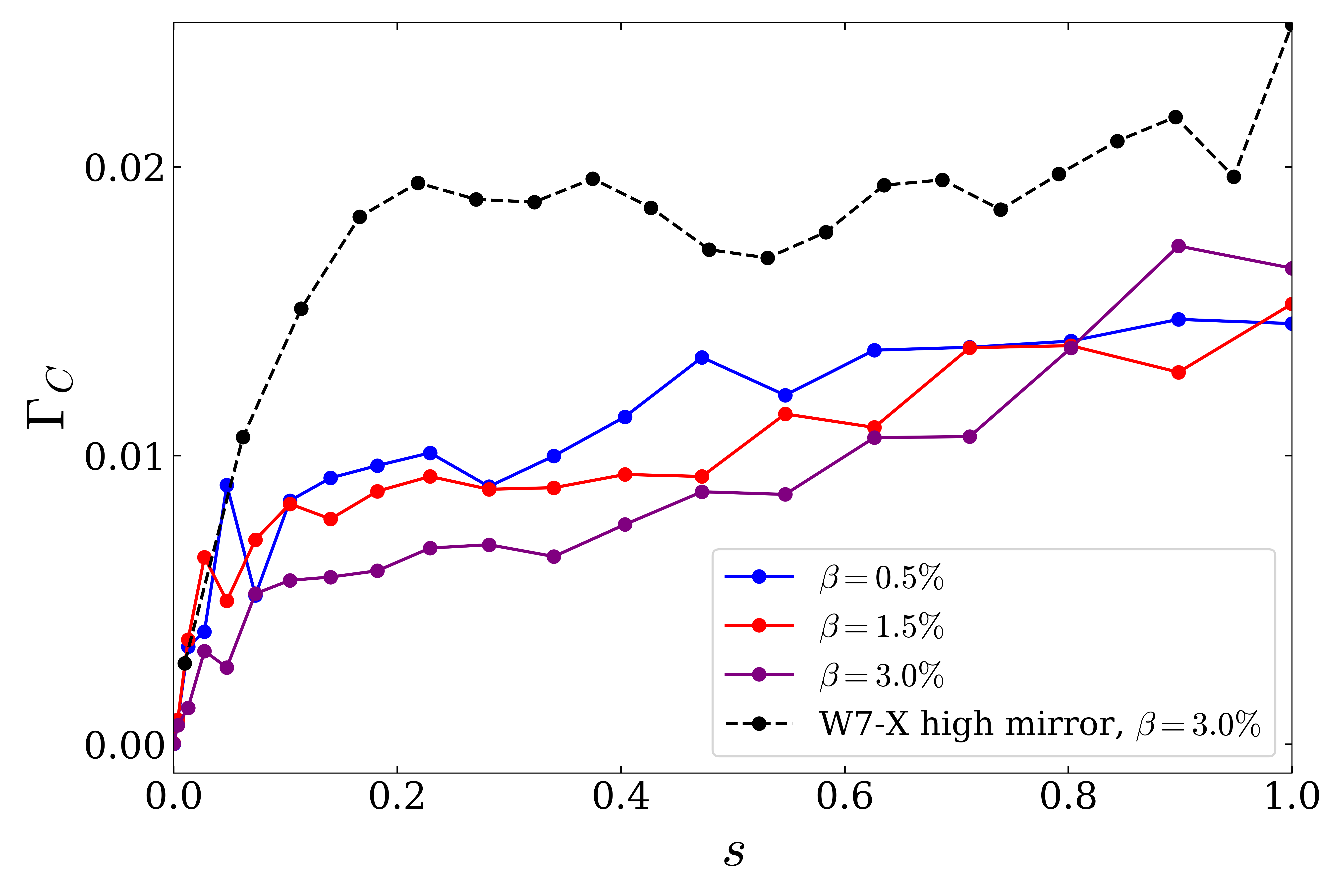}
\caption{Optimization of CIEMAT-pw1 with respect to the maximum-$J$ criterion: radial profiles of $B_{\mathrm{min}}$ and $B_{\mathrm{max}}$ (top) and of the $\Gamma_c$ figure of merit. \label{FIG_NCFI}}
\end{figure}

{\section{Results}\label{SEC_RES}}

Figure~\ref{FIG_EQ3D} presents two different views of the magnetic field strength on the $s=1$ flux surface of an optimized MHD equilibrium that we call CIEMAT-pw1. Its aspect ratio (the ratio between the major and minor radius) is 12.7, and its maximum elongation (at toroidal section $\phi=\pi/N_{fp}$, see figure~\ref{FIG_EQ3D}, top) is 6.8. 

\

{\subsection{CIEMAT-pw1, a magnetic configuration close to piecewise omnigenity}\label{SEC_BPWO}}

The level of piecewise omnigenity of CIEMAT-pw1 is assessed in figure~\ref{FIG_BPWO}: the top row compares $B(\theta,\zeta)$ at the radial position $s=0.5$ of the optimized equilibrium with $B_{pwO}(\theta,\zeta)$ for $p=2$ at the end of the optimization process. The relative difference is below 1\%, and it is of the order of 0.1\% for most of the flux surface. Additionally, the optimized equilibrium is compared with the $p\rightarrow\infty$ limit of $B_{pwO}(\theta,\zeta)$ (although this relative difference was not an optimization target, it is strictly speaking the quantity that determines closeness to piecewise omnigenity), and is shown to be larger, around 10\%, at specific regions of the flux surface. These values, which are larger than those obtained so far for omnigenous fields \cite{landreman2022preciseQS,goodman2023qi,dudt2023omni,liu2025omni}, could likely be made smaller by relaxing some of the other optimization criteria. Nevertheless, in a non-academic scenario, higher levels of piecewise omnigenity would likely be spoiled by finite $\beta$ effects or by error fields introduced by the coils (see section 2 of \cite{velasco2023flatmirror} for an omnigenous example). As predicted, figure \ref{FIG_BPWO} (bottom) shows a  degradation of the level of piecewise omnigenity with changes in $\beta$ and, specifically, with the distance to $s=0.5$. In particular, because of the relative importance of the $m=0$ modes, the magnetic field at the position closest to the axis, figure~\ref{FIG_BPWO} (bottom left), resembles the nearly-QI pwO field shown in figure \ref{FIG_EXAMPLES2} (bottom left).

We next show that this level of piecewise omnigenity is appropriate to ensure robustly good neoclassical transport properties for CIEMAT-pw1. Figure \ref{FIG_NC} shows the radial profile of the effective ripple, $\varepsilon_\mathrm{eff}$ (top), and the collisionality dependence of the $D_{31}$ bootstrap transport coefficient (bottom). These quantities \jl{(the latter computed with \texttt{MONKES}~\cite{escoto2024monkes})} encapsulate the dependence of neoclassical transport in the radial and parallel directions, respectively, on the details of the magnetic configuration~\cite{beidler2011ICNTS}. Comparison with W7-X shows that these quantities are significantly small, as expected~\cite{velasco2024pwO,calvo2025pwO}. The effective ripple is small at $s=0.5$, the optimized flux surface, for all $\beta$ values, and decreases even further towards the magnetic axis. It increases towards the edge, but always below that of the standard configuration of W7-X. The bootstrap transport coefficient is presented for the most adverse situation, zero radial electric field (where larger values are expected~\cite{beidler2011ICNTS,albert2025bootstrap}) and $s=0.06$ (far from $s_0$ and, additionally, where the collisionality is lowest for any realistic reactor scenario), and it displays values comparable to the high-mirror configuration of W7-X. \jl{The latter has a bootstrap current that, although larger than that of more recent QI reactor candidates, see e.g.~\cite{sanchez2026qi}, is considered small enough for a reactor}. 

Finally,  figure \ref{FIG_NCFI} (top) shows that, for sufficiently high $\beta$, both $\partial_s B_{\mathrm{min}}$ and  $\partial_s B_{\mathrm{max}}$ are positive up to $s=s_0$ in CIEMAT-pw1. This, as discussed in section~\ref{SEC_OPT}, will contribute to achieving the maximum-$J$ property (see also \cite{velasco2023flatmirror,rodriguez2024maxJ}). Indeed, the fast-ion proxy $\Gamma_c$ \cite{nemov2008gammac,velasco2021prompt}, which is in fact a phase-space average of $|\partial_\alpha J/\partial_s J|^2$, displays values that decrease with $\beta$ and that are significantly lower than for W7-X.

\

\subsection{Compatibility of CIEMAT-pw1 with reactor physics design criteria}\label{SEC_REACTOR}

A viable reactor candidate needs to satisfy a set of physics criteria that extend well beyond reduced bulk neoclassical transport, ranging from fundamental requirements to others that are considerably more advanced. Among the former, the magnetic configuration must be magnetohydrodynamically stable and possess good flux surfaces throughout the plasma volume. Among the latter, turbulent transport and alpha-particle losses must remain sufficiently low to reach and sustain the reactor operating point. Finally, the magnetic configuration must be compatible with the chosen divertor concept. These physics requirements were addressed during the optimization of CIEMAT-pw1 (see Appendix~\ref{SEC_DESC} for the specific set of optimization targets). In this section we analyze, for a range of plasma $\beta$, to what extent they were achieved. Where appropriate, this will be evaluated by comparison with the properties of W7-X, \jl{as it is relatively well established where its stellarator configuration stands in relation to a viable fusion reactor.}

\begin{figure}
\includegraphics[angle=0,width=0.95\columnwidth]{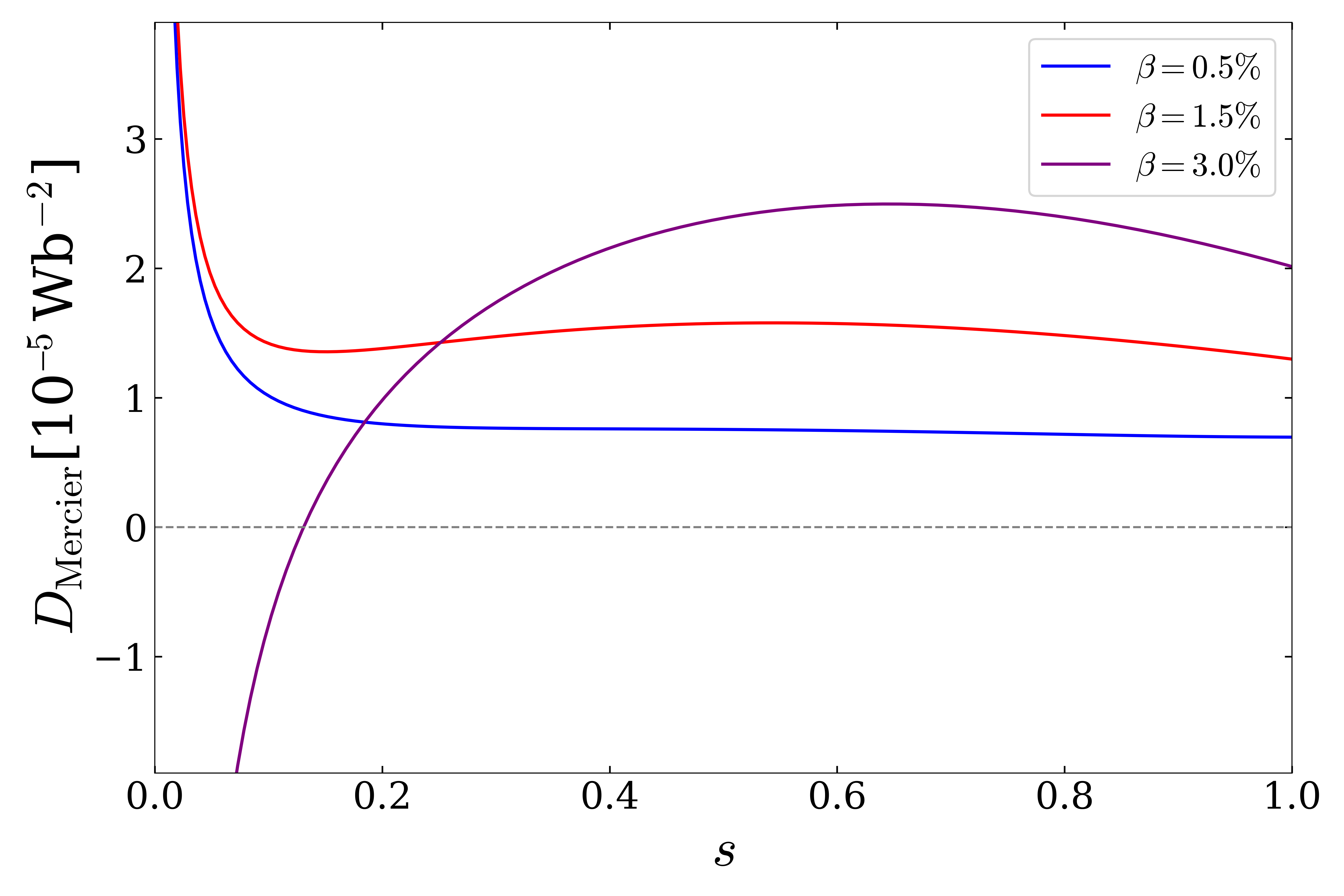}
\caption{Mercier stability criterion.\label{FIG_MHD}}
\end{figure}

The Mercier criterion is employed to assess MHD stabiity in CIEMAT-pw1, as shown in figure \ref{FIG_MHD}. The configuration is expected to be stable, since $D_{mercier}>0$ in most of the plasma radius (except at the highest $\beta$ values very close to the core, where simulations sometimes become unreliable). The increase of $D_{mercier}$ with $\beta$ is caused by the magnetic well, the main contributor to stability, which overcomes the destabilizing effect of the geodesic curvature. Ballooning stability has been confirmed by means of calculations with the code \texttt{COBRA}~\cite{sanchez2000boozerxform}, with negative growth rates for all the cases considered.

\begin{figure}
\includegraphics[angle=0,width=0.95\columnwidth]{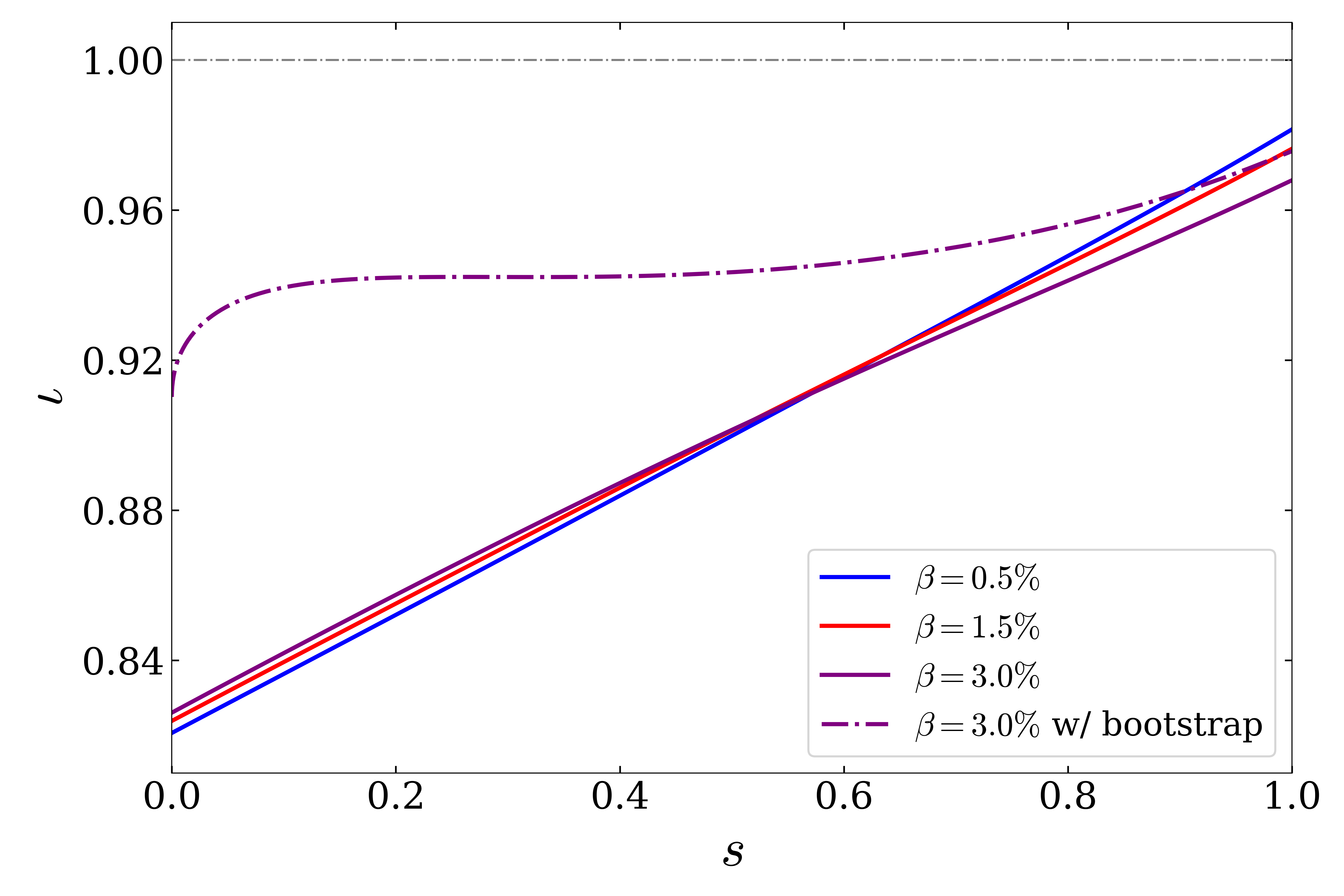}
\includegraphics[angle=0,width=0.95\columnwidth]{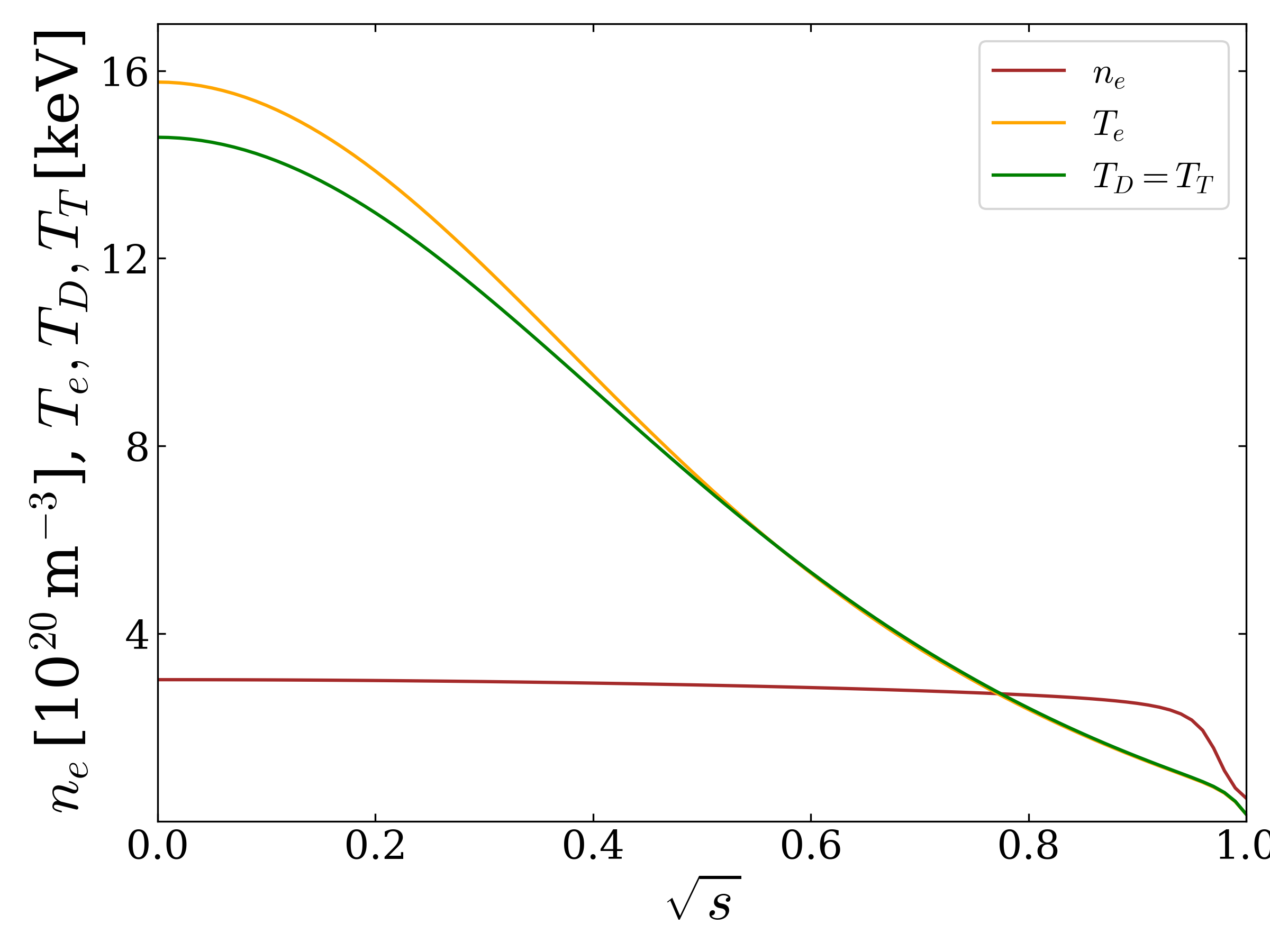}
\caption{Rotational transform: radial profile of $\iota$ (top) and plasma profiles of the reactor scenario employed for the bootstrap calculation (bottom).\label{FIG_IOTA}}
\end{figure}

The rotational transform profile of CIEMAT-pw1 is shown in figure~\ref{FIG_IOTA}. It has been designed with the general objective of avoiding low-order rational values in the core, as they could lead to the formation of magnetic islands~\cite{waelbroeck2009islands}, and to have the 5/5 rational value just outside the boundary, which could be the basis of an island divertor at $s>1$ ($s=1$ then would become the last closed flux surface). In the absence of a plasma current, $\iota$ grows approximately linearly with $s$ (with a similar slope to the one that achieved, for the particular case of CIEMAT-QI4X~\cite{sanchez2026qi}, an optimal balance between avoiding island overlap at the regions of $s<1$ where rational values could be prevented, and permitting a large magnetic island at $s>1$). In order to assess the effect of the bootstrap current on the rotational transform profile, a reactor physics design point is identified by scaling the magnetic configuration to a major radius of $R=18.5\,$m and a magnetic field strength on axis of $B_\mathrm{axis}=5.5\,$T (keeping $\beta=3$\%). Using the tools devised in~\cite{alonso2022reactor} and similar assumptions (in terms e.g. of fusion gain and energy confinement time scaling) the plasma profiles of figure~\ref{FIG_IOTA} (bottom) are obtained (quasineutrality between electron, deuterium and tritium species is imposed by setting $n_e=2n_D=2n_T$). Then, neoclassical transport is computed with \texttt{SFINCS} \cite{landreman2014sfincs} and the resulting bootstrap current is employed to recalculate the $\beta=3$\% equilibrium (given the uncertainties in the reactor plasma profiles, which would only be reduced by higher-fidelity modelling, e.g.~\cite{banonnavarro2023tango}, we do not attempt to obtain self-consistency, and rather limit ourselves to illustrating the size of the change in $\iota$ that is to be expected, along with its impact on confinement properties). Compatibility with an island divertor is supported by the fact that, in figure~\ref{FIG_IOTA} (top), the change in $\iota$ at the plasma edge is small. Close to the axis, where the collisionality is smaller, and the bootstrap current level higher, the change is larger, and goes in the direction of avoiding the 5/6 rational surface. We will confirm below that a change in $\iota$ of the size of that of figure~\ref{FIG_IOTA} (top) does not cause a large degradation of the transport properties.

Since the effective ripple is approximately one order of magnitude below that of W7-X (already considered sufficiently small for a reactor), turbulent transport is expected to be the main channel for energy loss of the bulk plasma in CIEMAT-pw1. In this work, it is computed with the gyrokinetic code \texttt{stella} \cite{barnes2019stella}, by means of flux-tube simulations with three kinetic species: deuterium, tritium and electrons.  Although electromagnetic effects will likely become important at the $\beta$ values of a reactor scenario, this type of simulations has been insufficiently validated by experiments in stellarators; we thus follow the example of \cite{goodman2024squids,regana2025qi,lion2025stellaris} and perform electrostatic simulations. Our results should then be considered as a baseline level for turbulent transport, which will then be likely modified (either reduced or increased) by the presence of electromagnetic effects~\cite{mulholland2023beta}, fast particles~\cite{disiena2020fi} and/or impurities~\cite{regana2024imp}, to name a few effects. For reference, reactor scenarios typically require turbulent heat fluxes $Q/Q_{i,gB}$ between $O(10^{-2})$ (at the core) and $O(10^{-1})$ (towards the edge)~\cite{alonso2022reactor}. Here, $Q$ is normalized by $Q_{i,gB}$, the gyro-Bohm ion transport level. Obtaining values in this range with our gyrokinetc simulations, as it will be the case, should then be considered a necessary, but not sufficient, condition for reactor relevance.

\begin{figure}
\includegraphics[angle=0,width=0.95\columnwidth]{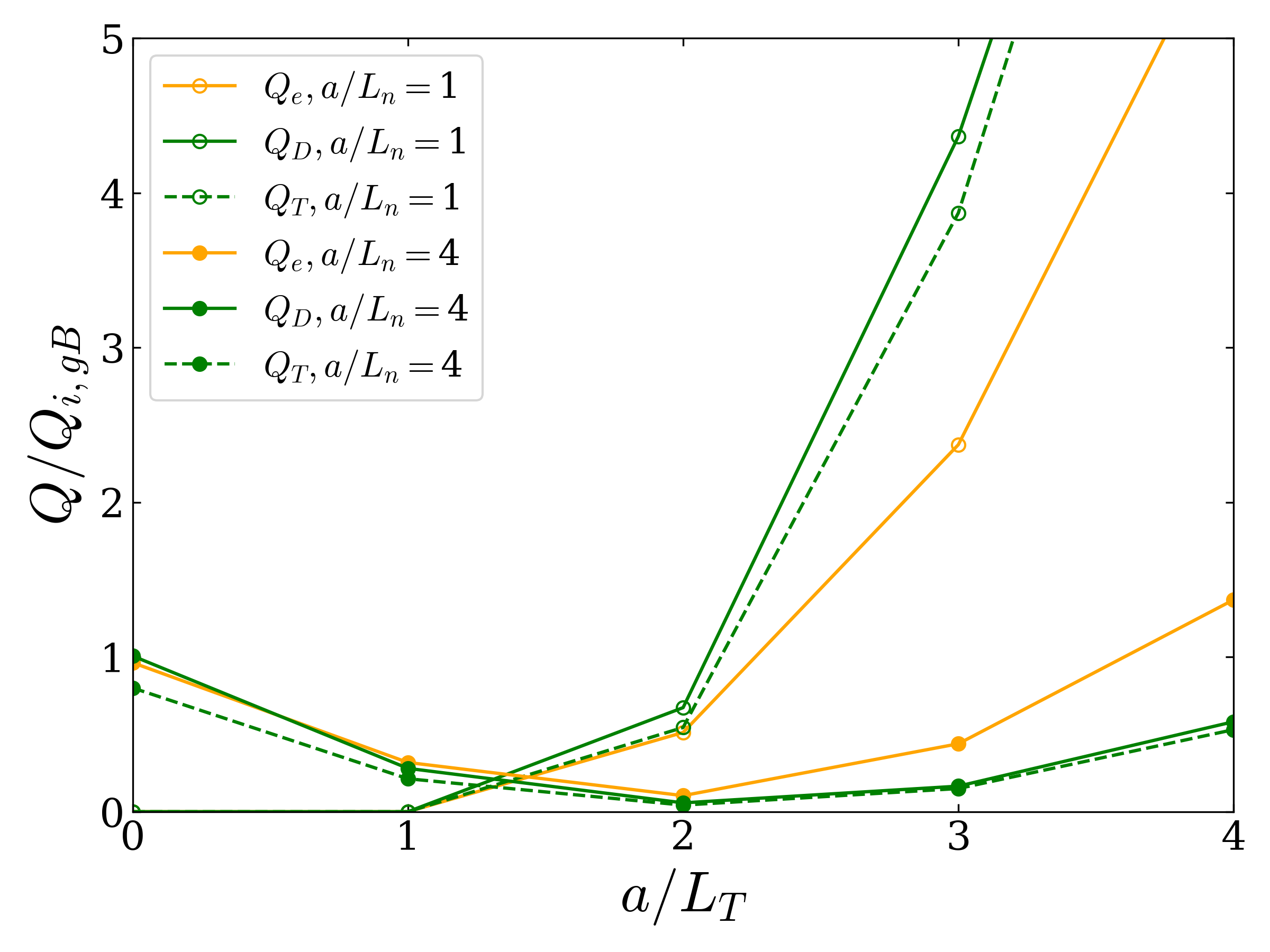}
\includegraphics[angle=0,width=0.95\columnwidth]{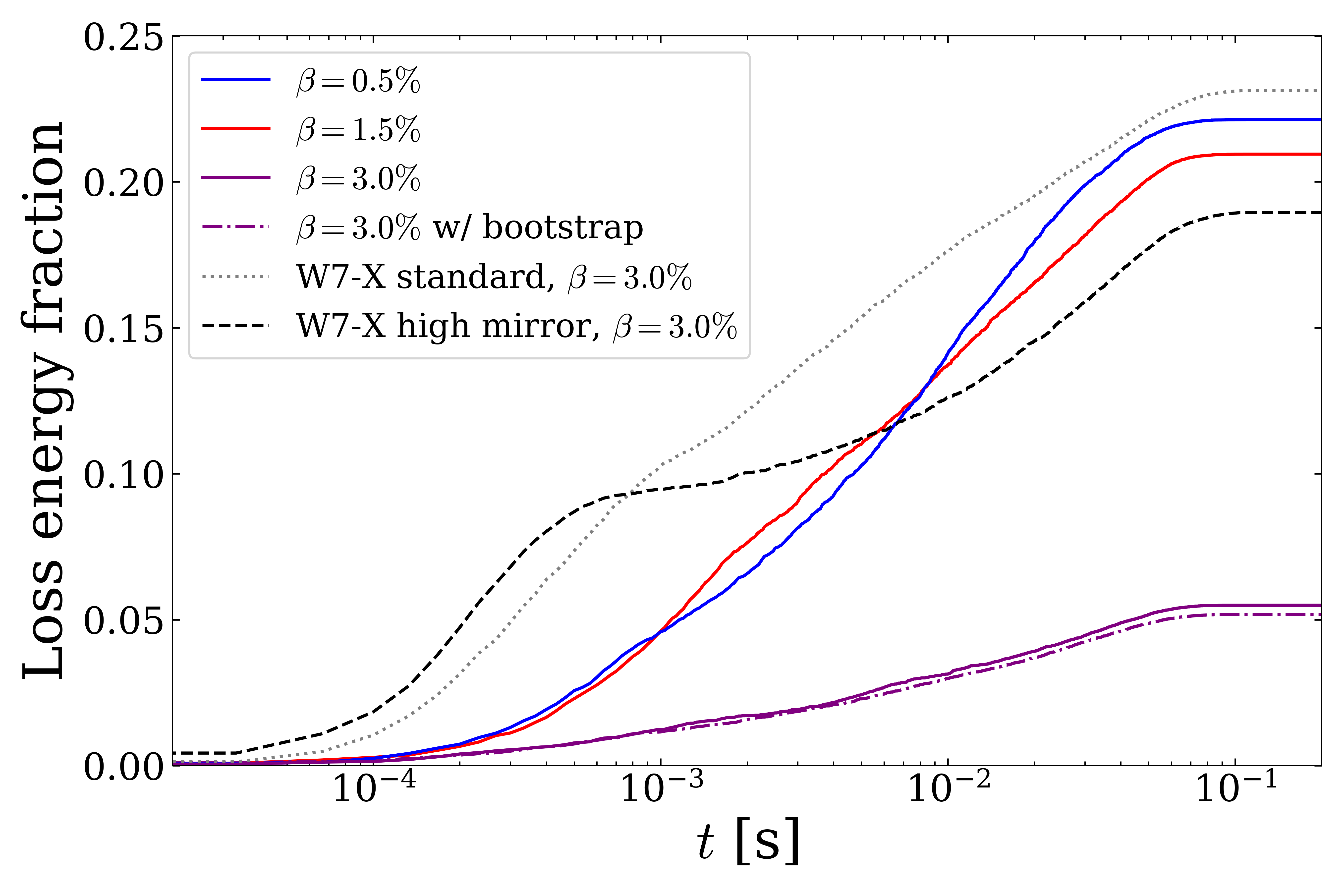}
\caption{Consequences of the optimization with respect to the maximum-$J$ criterion: turbulent fluxes of energy for the $\beta=3$\% configuration at $s=0.36$ as a function of the plasma gradients (top) and fraction of alpha particle energy lost through the boundary (bottom).}\label{FIG_MAXJ}
\end{figure}

Figure \ref{FIG_MAXJ} (top) shows the turbulent energy flux at $s=0.36$ as a function of the normalized temperature gradient, $a/L_{T}$,  for several species and values of the normalized density gradient, $a/L_n$, (in these simulations, $T_e=T_D=T_T$ and $n_e=2n_D=2n_T$). We observe that the deuterium and tritium fluxes are very close, and the electron flux has the same qualitative behaviour. By comparing the simulations for  $a/L_n=1$ and  $a/L_n=4$ at large $a/L_T$, we can confirm the estabilization expected from fulfilling, piecewisely, the maximum-$J$ property. On the other hand, for a small density gradient, a scenario that was not addressed in our optimization strategy, the energy fluxes grow rapidly with the temperature gradient. However, the values remain small (compatible with zero) up to a moderate value of $a/L_T$: following the procedure of~\cite{robertclark2026qi}, a critical gradient  $a/L_{T_\mathrm{crit}}$ between 1.5 and 2 can be roughly estimated. This value lies between that of W7-X and that of magnetic configurations specifically optimized for this aspect in the referenced work. Along the same line, when comparing the points with 
$a/L_n=1$ and $a/L_T=3$ to those from existing devices (ASDEX Upgrade, TJ-II, LHD, NCSX, and the W7-X standard configuration, see~\cite{thienpondt2025gradn}, figures 22 and 23), the configuration shows consistently similar or superior performance, at least for $\beta=3$\%. Altogether, these preliminary results indicate that pwO configurations are amenable to turbulence optimization.

Finally, the confinement of alpha particles of CIEMAT-pw1 is estimated with the \texttt{ASCOT} code~\cite{akaslompolo2019ascot} and shown in figure~\ref{FIG_MAXJ} (bottom). With an initial distribution that takes into account a realistic fusion reaction rate, the guiding-center trajectories of alpha particles are followed, including collisions with the background, until they have thermalized or have been lost through $s=1$ (in order to isolate configuration effects and to allow for a comparison with the literature, we employ for the reaction rate and for collisions of all cases the same profiles of density and temperature, taken from \cite{landreman2022bootstrap}). For $\beta=3\%$, the optimized equilibrium exhibits an alpha heating efficiency of 95\%, compatible with the assumptions of the reactor scenario. Moreover, even at low $\beta$, most of the lost alpha particles escape the plasma after they have transferred a fraction of their energy to the plasma (the slowing-down time is $\sim$ 0.1$\,$s), which limits the potential damage to plasma facing components \jl{(indeed, the $\Gamma_c$ proxy is considered a better predictor of prompt collisionless losses than of total losses~\cite{velasco2021prompt,paul2022fastions})}. Finally, we note that the fast ion confinement of the equilibrium, one of the transport properties most sensitive to details of the magnetic geometry, is not affected by the change in $\iota$ caused by the bootstrap current.


\section{Discussion}\label{SEC_SUMMARY}

Summarizing, in this work, we have radically advanced a new concept, piecewise omnigenity, for the confinement of magnetic fusion plasmas. Starting from a theoretical concept illustrated by a limited set of incomplete examples, we have demonstrated that piecewise omnigenity is a criterion that can actually be employed to design stellarator magnetic configurations that comply with the standard set of physics criteria required for a viable reactor candidate. The magnetic configuration CIEMAT-pw1 is presented in this work as a first example. The exploration of the configuration space of pwO fields has just been initiated, and better candidates, possibly combined with omnigenous fields~\cite{velasco2025parapwO}, will likely be found.

A key question, which has to be answered in the new line of investigation that this work opens, is to what extent pwO fields provide additional specific advantages with respect to omnigenous fields. To this respect, the resilience of the transport properties of CIEMAT-pw1 to $\iota$ changes presents an interesting example. From the physics points of view, mechanisms such as turbulence stabilization by magnetic shear (see e.g.~\cite{kessel1994shear,nadeem2001shear}) that are in principle not viable for QI or QS fields, could be so for pwO fields (in the sense of not being deleterious to neoclassical transport). \jl{Additionally, more compact configurations may become accessible by combining piecewise omnigenity with quasi-isodynamicity~\cite{velasco2026QIpwO,liu2026opwO} or quasi-axisymmetry~\cite{velasco2026QApwO}.} From the technological point of view, robustness against changes in the magnetic configuration is connected to the problem of coil complexity, a major driver of reactor cost. While this work focuses on physics properties and leaves technological details such as coils or compatibility with a breeder blanket for future work, Appendix \ref{SEC_COILS} shows some preliminary promising results, in line with partial results presented in~\cite{liu2025omni}. This is a very relevant matter that requires dedicated investigation.

\

\begin{acknowledgments}
The authors are grateful to EUROfusion’s TSVV12 team for its valuable feedback, to the \texttt{DESC} team, and to all the developers of the codes \texttt{VMEC}, \texttt{COBRA}, \texttt{MONKES}, \texttt{SFINCS}, \texttt{stella} and \texttt{ASCOT}. Calculations for this work made use of computational resources at Xula (CIEMAT), Turgalium (CETA-CIEMAT) and Marenostrum V (Barcelona Supercomputing Center). This work has been carried out within the framework of the EUROfusion Consortium, funded by the European Union via the Euratom Research and Training Programme (Grant Agreement No 101052200 EUROfusion). Views and opinions expressed are however those of the author(s) only and do not necessarily reflect those of the European Union or the European Commission. Neither the European Union nor the European Commission can be held responsible for them. This research was supported by grants PID2021-123175NB-I00 and PID2024-155558OB-I00, Ministerio de Ciencia, Innovaci\'on y Universidades, Spain. This work was conducted as part of the FUSION-EP Master’s Programme, and V.F.-P. gratefully acknowledges the programme for its support.
\end{acknowledgments}

\section*{Author contributions}

Conceptualization: V.F.-P., J.L.V., E.S., I.C..
Data curation: V.F.-P., J.L.V., E.S., J.M.G.-R..
Formal analysis: V.F.-P., J.L.V., E.S., J.M.G.-R., J.A.A., D.C..
Funding acquisition: J.L.V., J.M.G.-R., I.C..
Investigation: V.F.-P., J.L.V., E.S., R.G., J.M.G.-R., J.A.A., D.C..
Methodology: V.F.-P., J.L.V., E.S., R.G., J.M.G.-R., J.A.A., I.C., D.C.
Project administration: V.F.-P., J.L.V., E.S..
Software: V.F.-P., R.G..
 Supervision: J.L.V., E.S..
Validation: J.L.V., E.S..
Visualization: V.F.-P., J.L.V., E.S.. 
Writing – original draft: J.L.V., V.F.-P., E.S..
Writing – review \& editing: R.G., J.M.G.-R., J.A.A., I.C., D.C..

\section*{Data availability}

The data supporting the findings of this article are available at https://zenodo.org/records/20390611.

\appendix

\section{Optimization \jl{and simulation} details}\label{SEC_DESC}

\begin{table*}
\centering
\begin{tabular}{c l c c c}
 \hline
Objective & Type & Target/bounds & Purpose \\ [0.5ex] 
 \hline\hline
Force balance residual & Volumetric &  $0$ & Accurate MHD equilibrium \\
Toroidal current & Profile & $0$ & Stellarator constraint \\
Major radius & Scalar & same as initial configuration & No trivial optimization \\ 
$B$ on axis & Scalar & same as initial configuration & No trivial optimization \\
2nd principle curvature of boundary & Set of scalars &  $> (-100, 10)$ (m$^{-1}$) & Simple boundary shape \\
Maximum elongation & Scalar  & 7 & Simple boundary shape \\
Rotational transform &Profile &  $\iota=0.82+0.16s$ (w/o bootstrap)  & Good flux surfaces,  \\
 &  &   & compatibility with island divertor \\

Mercier criterion &Profile &  $>10^{-4}\,$Wb  & MHD stability \\
Mirror ratio  & Scalar &  $<$ 0.15 & Reduced fraction of trapped particles \\
$(B_{\mathrm{max}}-B_{\mathrm{min}})/(B_{\mathrm{max}}+B_{\mathrm{min}})$ at $s=1$ & & & (potential drive of transport) \\
Deviation from $B_{pwO}$ at $s=0.5$ &Surface &  $0$ & Collisionless confinement of orbits \\
 \hline
\end{tabular}
\caption{Objectives used for optimization with their target/bounds and purpose. }
\label{TAB_DESC}
\end{table*}

Table~\ref{TAB_DESC} provides a list of the different targets employed in the optimization process with \texttt{DESC}, along with some motivation (precise definitions of the targets as implemented in \texttt{DESC} can be found in a Zenodo repository~\cite{zenodo}). \jl{We note that the general motivation for the $\iota$ target is given in section \ref{SEC_REACTOR}, but similar $\iota(s)$ functions would have likely yielded similarly good results.} \jl{The aspect ratio was monitored but not included as an explicit optimization target.} As initial condition, the boundary of configuration A of~\cite{bindel2023direct} was changed to $N_{fp}=5$. \jl{The Zenodo repository~\cite{zenodo} also includes scripts and inputs that allow to reproduce the main calculations of this work with the appropriate resolution.}

\begin{figure}
\includegraphics[angle=0,width=0.95\columnwidth]{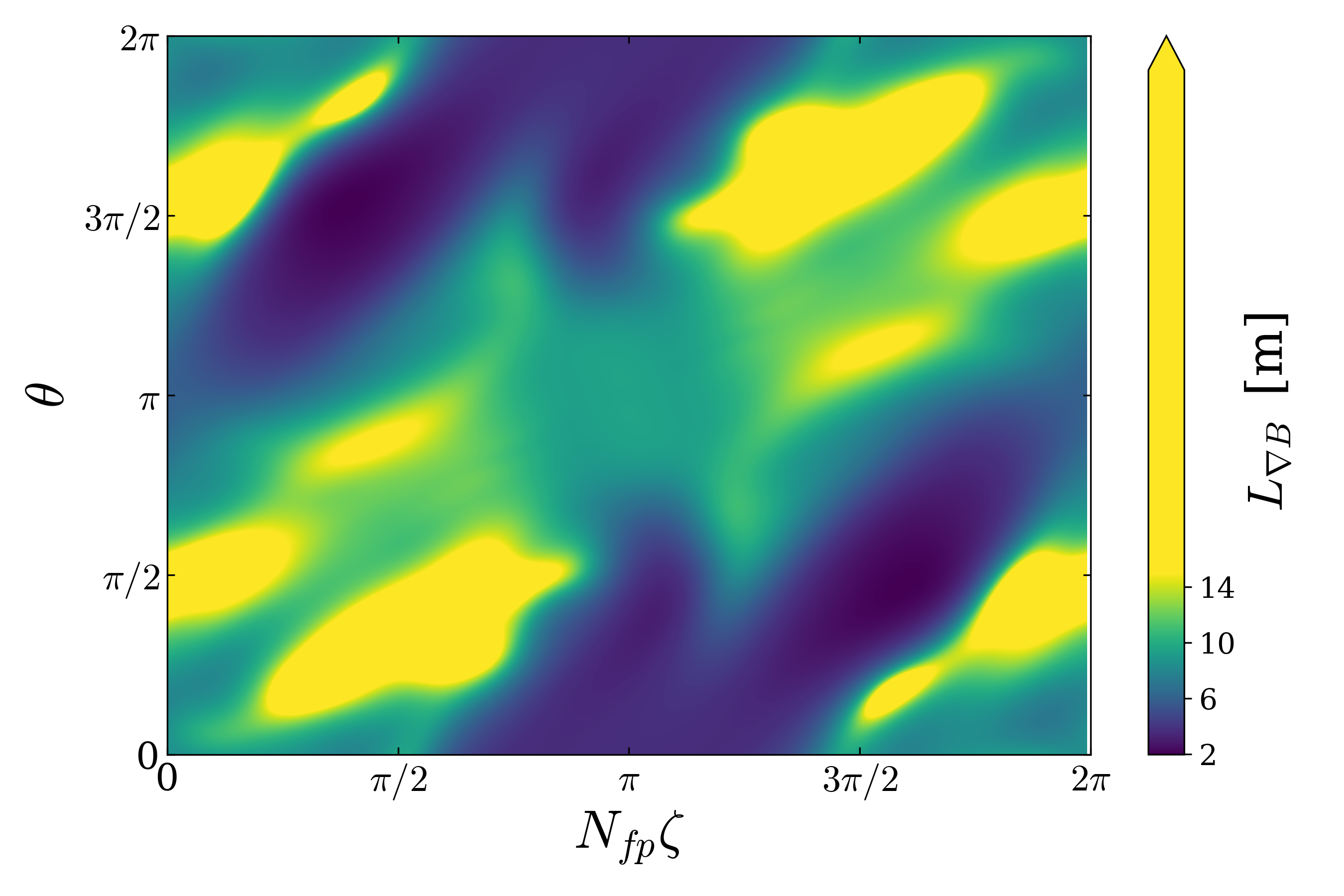}
\caption{Estimate of coil complexity. \label{FIG_LGRADB}}
\end{figure}

\section{Coil feasibility}\label{SEC_COILS}

Once the physical properties of the equilibrium have been analyzed, the next step would be to look for a set of magnetic coils that are able to reproduce the obtained magnetic field without compromising its favourable properties \jl{(this would in turn allow for a detailed design of the edge island structure, so that it can be employed for heat and particle exhaust). This is usually considered an independent optimization problem, see e.g.~\cite{gil2026coils,veksler2026island},} and will not be addressed in this work. However, we will use a proxy, the magnetic gradient scale length $L_{\nabla B}$~\cite{kappel2024gradient}, to obtain a first, albeit rough, indication of coil feasibility. This quantity provides an estimate of the minimum distance at which a set of reasonably simple modular coils should be placed in order to reproduce correctly the magnetic configuration. The criterion of magnetic field fidelity tends to push the coils to be close to the plasma (otherwise, they may become very complicated if they need to generate high harmonics of $B$). On the other hand, in a reactor, the plasma-coil distance must exceed a certain threshold in order to have space for the neutron shielding required by the coils and for a breeding blanket able to produce tritium at an adequate rate (these considerations would naturally be less pressing for a non-nuclear experimental device). A compromise needs to be found.

Figure \ref{FIG_LGRADB} shows $L_{\nabla B}$ at flux surface $s=1$ of the magnetic configuration. From these results, the minimum plasma-to-coil distance can be estimated to be approximately 1.3$\,$m (details of the model, which should be considered a rough estimate, are given in~\cite{kappel2024gradient}). This distance may be somewhat tight to accommodate the required components in a reactor. However, the predicted distance is much larger in other regions, which might compensate for this and provide an adequate tritium breeding ratio. An assessment of this requires the design of coils and neutronics analyses that are left for future work.

%

\end{document}